\documentclass{aa}
\usepackage{psfig,graphics}

\def\mass#1{${\mathrm{#1\, M}_\odot}$}

\def\simgr{\mathbin{\;\raise1pt\hbox{$>$}\kern-8pt\lower3pt\hbox{$\sim$}\;}}
\def\simlr{\mathbin{\;\raise1pt\hbox{$<$}\kern-8pt\lower3pt\hbox{$\sim$}\;}}

\begin{document}


\thesaurus{06         
stars
           (08.16.4;  
           08.06.3;   
           08.11.1;   
           08.16.1;   
           08.19.1)   
          }

\title{Long Period Variable Stars: galactic populations and infrared 
luminosity calibrations.}

\author{M.O. Mennessier\inst{1}
        \and N. Mowlavi\inst{2}
        \and R. Alvarez\inst{3}
  	\and X. Luri\inst{4}
        }

\offprints{ M.O. Mennessier }
 
\institute{ Universit\'e de Montpellier II and CNRS, 
	    G.R.A.A.L., cc072,
            F--34095 Montpellier CEDEX 5, France
           \and 
	    Observatoire de Gen\`eve, 
	    CH--1290 Versoix, Switzerland
           \and	
            Institut d'Astronomie et d'Astrophysique,
            Universit\'e Libre de Bruxelles, C.P. 226,
            B--1050 Bruxelles, Belgium
           \and	
            Departament d'Astronomia i Meteorologia,
            Universitat de Barcelona,
            Avda. Diagonal 647,
            E08028, Barcelona, Spain        
          }

\date{Revised 21 Dec.2000 /Accepted 16 May 2001}

\titlerunning{Photometric and kinematic properties of LPVs}
\authorrunning{M.O. Mennessier et al.}

\maketitle 


\begin{abstract}
		
In this paper HIPPARCOS astrometric and kinematic data are used to
calibrate both infrared luminosities and kinematical parameters of Long
Period Variable stars (LPVs). Individual absolute K and IRAS 12 and 25
luminosities of 800 LPVs are determined and made available in electronic
form.

The estimated mean kinematics is analyzed in terms of galactic
populations.  
LPVs are found to belong to galactic populations ranging from the thin
disk
to the extended disk.  An age range and a lower limit of the initial
mass
is given for stars of each population.  A difference of $1.3$mag in K
for
the upper limit of the Asymptotic Giant Branch is found between the disk
and old disk galactic populations, confirming its dependence on the mass
in
the main sequence.

LPVs with a thin envelope are distinguished using the estimated mean
IRAS
luminosities. The level of attraction (in the classification sense) of
each
group for the usual classifying parameters of LPVs (variability and
spectral types) is examined.


\end{abstract}


\section{Introduction} \label{sec_intro}

Long period variables (LPV) form an important class of red giant stars.
They show more or less regular photometric variability with amplitudes
reaching up to 8 magnitudes and periods up to 600 days. They
traditionally
comprise Miras, semi-regular (SR) and irregular (L) variables according
to
the amplitude and the regularity of their visual light curves.  They are
known to be either O-rich or C-rich, and comprise thus M, S and C stars.
More recently OH-IR sources have been found from infrared and radio
observations
showing that they belong to the LPV population with periods up to 2000
days. Those sources emit in the infrared and radio wavelengths, and are
not
associated with any detectable counterpart in optical wavelengths.

The brightest LPVs are luminous enough to be observed at long distances,
providing information on the host galaxy, like the Magellanic Clouds
(see
Van Loon et al., 1999 as an example). While the ranges of masses and
ages
of LPVs are still the subject of discussion, it is generally accepted
that they
are large, and are therefore considered as very good tracers of 
galactic history.

The determination of the characteristics of individual LPVs is usually a
delicate task due to the complexity of the dynamic and chemical
phenomena
to be considered. A statistical study using all available data of a
large sample of LPVs is often needed. A rough example of such an
approach could
be the relation between the mean visual light curves and the infrared
colors of C and O-rich LPVs already presented in Mennessier et al.
(1997a).
In this paper, HIPPARCOS astrometric data and the available
multi-wavelength (K, IRAS 12 and 25) infrared photometric measurements
allow us to calibrate multi-wavelength luminosities and to discriminate
between different galactic populations -- and thus different ranges of
initial
masses (${\cal M}_{ms}$) -- among the LPVs according to their
kinematical
properties.  In a second step, individual K and IRAS absolute magnitudes
are estimated for all the 800 considered LPVs using a powerful
statistical
estimator.\\

Our sample of LPV stars and the data used are described in Sect.
\ref{Sect:sample}. The statistical method specifically developed for the
study of HIPPARCOS samples is summarized in Sect.~\ref{Sect:statistics}.
Sect. \ref{Sect:groups} presents the discriminated groups of LPVs
resulting
from our statistical analysis, while Sect. \ref{Sect:individual}
analyzes
the results derived for individual stars. Finally, Sect.
\ref{sec_propgrups} reviews the crossed properties derived from the
analysis at different wavelengths.

\section{The sample of LPV stars} \label{Sect:sample}

\subsection{Data} \label{Sect:data}

In order to benefit from the accurate astrometric data made available by
the HIPPARCOS satellite, we use in our study the sample of all LPV stars
observed on this mission, i.e. the LPVs brighter than 12.5 mag in V
during
more than 80\% of their variability cycle. The sample is composed of
about
900 stars which are either of type M (O-rich), C (C-rich) or S
(O/C$\simlr$1). They include Mira, SR (of both type {\sl a} and {\sl b})
and L variables.

Astrometric data is taken exclusively from the HIPPARCOS Catalogue
(Perryman et al. 1997) to provide a homogeneous data set. Radial
velocities
are taken from the HIPPARCOS Input Catalogue (HIC; Turon et al. 1992).

Photometric data are gathered from various sources. V magnitudes
($m_{V}$)
are taken from the HIC. They correspond to the magnitudes given in the
General Catalogue of Variable Stars (GCVS; Kholopov et al. 1985),
corrected
as described in the HIC volumes to obtain mean magnitudes at the maxima
of
light. K magnitudes ($m_{K}$) are taken from the Catalogue of Infrared
Observations (Gezari et al. 1996), and include the large set of JHKL
measurements of LPVs by Catchpole et al. (1979) and the measurements by
Fouqu\'e et al. (1992),  Guglielmo et al. (1993), Groenewegen et al.
(1993), Whitelock et al. (1994), Fluks et al. (1994), Kerschbaum \& Hron
(1994), Kerschbaum (1995) and Kerschbaum et al. (1996). Infrared
magnitudes
are derived from the $F_{12}$ and $F_{25}$ fluxes measured at 12 and 25
micrometers respectively by the infrared astronomy satellite (IRAS). We
use 

\begin{equation} 
  \label{Eq:12} 
  m_{12} = 3.63 - 2.5\times \log F_{12}
\end{equation} 

\noindent and 

\begin{equation} 
  \label{Eq:25} m_{25} = 2.07 -2.5\times \log F_{25}, 
\end{equation} 

as given in the IRAS-PSC catalog (vol.1, Sect.~VI.C.2). One should note
that not all authors use this definition for the infrared magnitudes
$m_{12}$ and $m_{25}$. The color index $m_{25}-m_{12}$ used by Van der
Veen
and Habing (1988), for instance, is higher than the one deduced from
Eqs.~\ref{Eq:12} and \ref{Eq:25} by $1.56$mag.

Among the $\sim$900 stars of our sample, the number of stars for which
V,
K and IRAS infrared magnitudes are available amounts to 882, 652 and
793,
respectively, with 608 stars having both K and IRAS magnitudes.

Finally, variability and spectral types are taken from the GCVS.

\subsection{Selection effects} \label{Sect:selection effects}

The main selection bias in our sample comes from the HIPPARCOS magnitude
limit $V<12.5$~mag (see Sect.~\ref{Sect:data}). This selection is well
determined and thus easy to take into account in the statistical
analysis.

\begin{figure}
\centerline{\psfig{figure=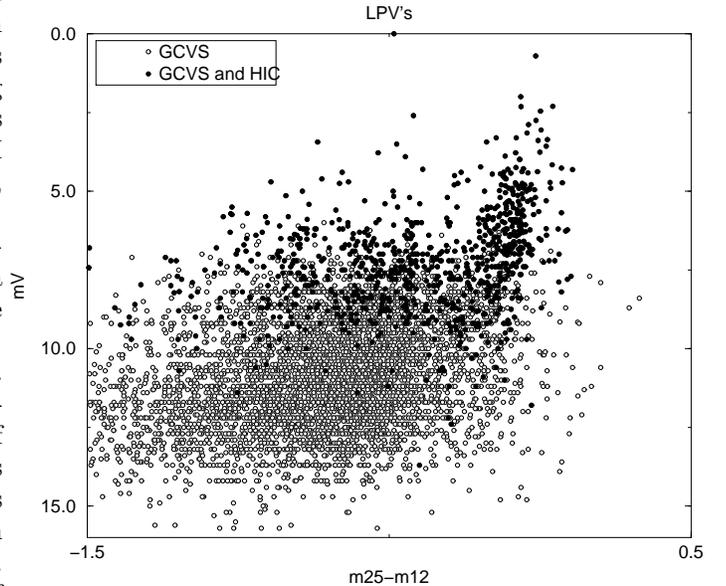 ,height=9cm ,angle=-90}}
\caption{Distribution of the GCVS LPVs according to apparent visual
magnitude at 
         maximum luminosity and IRAS color index. Stars observed
         by HIPPARCOS are indicated}
\label{fig_biaisobs}
\end{figure}

The characteristics of LPVs cause another bias related to the magnitude
limit of the sample. LPV stars are evolved red giants, often
characterized
by the formation of dust around them. The presence of a dusty
circumstellar
envelope affects the stellar spectrum by reducing their visible light.
As a
result, obscured LPVs are under-represented in our sample, because the
HIPPARCOS selection was done on the basis of the visible magnitude. The
importance of this bias can be estimated by comparing the number of
stars
included in the HIC with the number recorded in the GCVS. This
comparison
is shown in Fig.~\ref{fig_biaisobs} as a function of the V magnitude
$m_V$
and the color index $m_{25}-m_{12}$, where we consider all LPVs of the
GCVS
for which either the visual ($m_V$) or the photographic ($m_P$)
magnitude
at maximum is given, and assuming $m_P - m_V =1.8$ as the mean value for
LPVs. All stars from the HIC, represented by filled circles in
Fig.~\ref{fig_biaisobs}, are found to have $m_V > 12.5$, as expected
(the
very few exceptions being most probably due to the fact that the assumed
$m_P - m_V = 1.8$ relation does not apply to them).
Fig.~\ref{fig_biaisobs}
also shows that the number of stars included in the HIC (relative to the
number of stars recorded in the GCVS, represented by filled and open
circles in Fig.~\ref{fig_biaisobs}) decreases with increasing
circumstellar
envelope thickness (i.e. decreasing $m_{25}-m_{12}$ index). This bias is
further discussed in Sect.~\ref{sec_envbias}.

It must be noted that the GCVS itself is, of course, not exhaustive, and
is
certainly biased at the expense of the reddest stars. OH-IR stars, for
example, are not well represented in the GCVS sample. For these reasons,
a
statistical method which can take into account all these biases is
necessary for our analysis. This method is described in the next
section.

\section{The statistical method} \label{Sect:statistics}

In this paper the Luri-Mennessier (LM) statistical method, described in
Luri et al. (1996), is used to analyze the sample of LPV stars.  The
method
has been specifically designed to exploit the HIPPARCOS data and thus is
suitable for our purposes. This method has already given fruitful
results, in
particular for Barium stars (Mennessier et al., 1997c), Ap-Bp stars
(Gomez
et al., 1998) and the LMC distance modulus (Luri et al., 1998).  

The use of appropriate statistical methods for the exploitation of 
the HIPPARCOS astrometric data is crucial in order to obtain correct
results.
Otherwise the values obtained can be affected by strong biases and the
precision of the data will not be fully used. A discussion on the
correct use of HIPPARCOS data and recommendations on analysis techniques
can be found in Brown et al. (1997). 

The LM method used in this paper is especially appropriate for use with
the 
HIPPARCOS data. We refer to Luri et al. (1996) for a detailed 
description, and only briefly summarize here some of its main features.

First of all, the stellar {\sl population} from which the sample is
extracted is assumed to be composed of several distinct {\sl groups}.
These
groups can differ in  kinematics, luminosity or spatial distribution
and
its number is {\it a priori} not known. Therefore, using a sample
extracted from
this base {\it population} and taking into account the selection
criteria
used to create it, the LM method:

\begin{itemize}

\item  determines the number of significant discriminating groups and
the 
       percentage of each of them in the {\it population};

\item  produces, for each group, unbiased estimates 
       of the parameters of the model used to describe it, i.e.:

  \begin{itemize}
    \item a Schwarzschild ellipsoid velocity distribution characterized
by
          ($U_0, V_0, W_0, \sigma_U, \sigma_V, \sigma_W$).
          $U_0, V_0$ and $W_0$ point towards
          the galactic center, the galactic rotation, and the north
          galactic pole, respectively;
    \item an exponential distribution of the number of stars in the
          direction perpendicular to the galactic plane, with a  scale
          height $Z_0$;
    \item a normal distribution of the absolute magnitude in 
          the bandpass of the used observed magnitudes characterized by
the
          parameters $M_0$ and $\sigma_M$;
   \end{itemize}

\end{itemize}

The results of this first step for our LPVs sample are given in Sect.

The minimum input data needed by the LM method are the measured
positions,
proper motions and apparent magnitudes of the stars, but it can also use
the parallaxes and radial velocities if available. The method takes into
account the selection effects of the sample, the observational errors,
the
galactic rotation and the interstellar absorption.

In a second step, once the groups are identified and parametrized, the
method:

\begin{itemize}

\item computes for each star of the {\it sample}, 
      the {\it a posteriori} probability that the star 
      belongs to a given group;

\item uses a Bayesian rule to assign each star to a group;

\item uses a statistical estimator to obtain  estimations of individual
      distances and luminosities for each star.

\end{itemize}

This second step is presented in Sect.~\ref{Sect:individual} where the
representativity of our sample with respect to the kinematic and
photometric properties of the population is also discussed.

\section{Calibrations of the LPVs population} \label{Sect:groups}

The LM method was applied four times, as described in
Sect.~\ref{Sect:statistics}, once for each photometric bandpass:  V --
results already presented in Mennessier et al. (1997b) --, K, 12 and
25 -- in the present paper. In principle, one could assign a
joint luminosity distribution to two or more bandpass magnitudes
simultaneously, and the LM method would separate the sample into
stellar groups consistent with all the photometric measurements
together. This option, however, requires a perfectly well known
relationship between the different magnitudes in order to define a
joint distribution function as realistic as possible for all the band
passes. The correlation between the near-infrared (K) and IRAS
infrared properties presently cannot be well modeled and 
very likely has a non-unique form depending on the stellar and
circumstellar evolutive stage along the AGB. Thus, we decided not to
couple the photometric band passes and to calibrate each luminosity
separately.  Furthermore, bandpasses are related to different physical
processes and can provide separate interesting information:  V is
greatly affected by absorption molecular lines, K reflects the stellar
emission, and IRAS bandpasses depend on the nature and density of
grains in the circumstellar envelope.

The LM method is simultaneously sensitive to kinematics and luminosity
and thus the number of significant discriminating groups depends on
both these characteristics and is not necessarily the same for the
different bandpass analyses. Furthermore, the samples used are not the
same, and this can also affect the number of discriminated groups.

Six distinct groups are identified in the V magnitudes, three in K and
four in each of the two IRAS magnitudes. Those are successively
analyzed in terms of the classical galactic populations. Although the
number of groups is found to be different for each analysis, the
groups present similarities in their kinematical composition and with
respect to the galactic populations (see sect. \ref{sec_propgrups}).

\subsection{The V band} \label{Sect:V groups}

An analysis of the six groups identified in the V band has been 
presented in Mennessier et al. (1997b).  In order to compare these
results with the ones obtained for infrared calibrations, the main
results are summarized here.  Table~\ref{tab_estiV} reviews 
the estimated mean parameters for the analysis corresponding to the 
V luminosity at the phase of maximum light.  The LPVs are found to belong 
to all galactic populations from disk to very extended disk. We wish to 
emphasize three points:

\begin{itemize}

\item LPVs belonging to the bright disk population (BD) have
     a mean luminosity of $M_V$=-3.6 and $Z_0$=104pc, while
     the corresponding values 
     for the disk population (D) are $M_V$=-1.0 and $Z_0$=126pc.
     Among LPVs belonging to BD population there are probably stars 
     at the upper limit of the AGB or even at the first step of the
     post-AGB state.

\item the main group (44$\%$) has an estimated  scale height of 
     $Z_0$=249pc. Its kinematics is typical of the old disk population.

\item a few LPVs (less than 2$\%$) belong to the extreme 
     extended disk (ED), with $Z_0$ greater than 1200pc and a very
     large Schwarzschild ellipsoid 
     velocity distribution.  This most probably indicates a birth of those 
     stars early in the evolution of our Galaxy. Those stars should thus 
     be a metal-deficient disk population (see \ref {sec_age}).
     This is consistent with the bright luminosity found ($M_V$=-2.8). 

\end{itemize}

\begin{table}
\caption{V calibration: estimated parameters of the different groups and 
	   percentage of the sample into each of them.
        }
\label{tab_estiV}
\begin{center}
\begin{tabular}[h]{|l|cccccc|}
      \hline
            &        &       &      &     &    &    \\[-3pt]
      Group & BD     &  D    &  OD1 & OD2 & TD & ED \\[+5pt]
      \hline
                     &      &      &      &       &       &       
\\[-3pt]
      $M_V$          & -3.6 & -1.0 & -1.2 &  -0.2 &  -1.2 &   -2.8 \\
      $\sigma_{M_V}$ &  1.4 &  0.8 &  0.2 &   1.0 &   0.5 &    1.2 \\
      $U_0$          &  -10 &   -6 &  -44 &    -1 &   -34 &   -61 \\
      $V_0$          &  -11 &   -6 &  -35 &   -21 &   -84 & -235 \\
      $W_0$          &  -13 &   -6 &   -6 &   -10 &   -19 &  -20 \\
      $\sigma_{U_0}$ &   13 &   24 &   28 &   37  &    77 &  188 \\  
      $\sigma_{V_0}$ &   14 &   14 &   25 &   23  &    29 &  126 \\ 
      $\sigma_{W_0}$ &    9 &    9 &   22 &    23 &    65 &     72 \\
      $Z_0$          &  104 &  126 &  217 &   249 &   409 &   1227 \\
      $\%$           &    8 &   25 &   13 &    44 &     8 &    2  \\
      \hline
\end{tabular}
\end{center}
\end{table}

\subsection{The K band} \label{Sect:K groups}

Only three groups are identified in the K band. From their kinematics
and spatial distribution, given in Table~\ref{tab_estiK}, they can be
interpreted as the galactic disk (D), old disk (OD) and extended disk
(ED) populations. They are similar to the four main groups identified
in the V band (Sect.~\ref{Sect:V groups}), except that the disk and a
part of the old disk population seem to be mixed.

In a previous analysis of a sample restricted to O-rich Miras (Alvarez
et al.,1997), only two groups were found.  One corresponded to the
extended disk population, with a percentage of 17, in agreement
with our result i.e. 18/142=13$\%$ of O-rich Miras belonging to the ED 
group (see table~\ref{tab_contengency}). The other group mixed disk and old
disk populations. In the present paper, a more numerous sample allows a
more refined separation of the kinematic populations.

\begin{table}
\caption{K calibration: estimated parameters of the different groups  
	   and percentage of the LPVs population in each of them.
        }
\label{tab_estiK}
\begin{center}
\begin{tabular}{|l||rr|rr|rr|}
\hline
        & \multicolumn{2}{c}{Group D} & \multicolumn{2}{c}{Group OD} 
        & \multicolumn{2}{c}{Group ED} \\
        & est. & $\sigma$              & est. & $\sigma$             
        & est. & $\sigma$ \\
  \hline
   \hline
$K_0$                   &  -6.1 & 0.4 &  -6.0 & 0.7 &   -5.3 & 0.8 \\
$\sigma_K$              &   1.1 & 0.3 &   0.7 & 0.4 &   1.4 & 0.5 \\
$U_0$                   &  -7 & 10 &  -17 & 27 &   -21 & 14 \\
$\sigma_U$              &   29 & 9 &  45 & 16 &  111 & 11 \\
$V_0$                   &  -12 & 8 & -36 & 25 &  -123 & 12 \\
$\sigma_V$              &  16 & 5 &  27 & 11 &  69 & 18  \\
$W_0$                   &  -9 & 6 & -6 & 9 &  -20 & 19 \\
$\sigma_W$              &  12 & 3 &  26 & 13 &  90 & 25  \\
$Z_0$                   &  184 & 44 & 268 & 85 &  782 & 313 \\
$\%$                    &  60 &   & 35 &  &  5 &  \\ 
 \hline
 \end{tabular}
 \end{center}
\end{table}

\begin{table}
\caption{12 calibration: estimated parameters of the different groups
and 
	   percentage of the LPVs population in each of them
        }
\label{tab_esti12}
\begin{center}
\begin{tabular}{|l||rr|rr|rr|rr|}
\hline
        & \multicolumn{2}{c}{Group D} & \multicolumn{2}{c}{Group ODb} 
        & \multicolumn{2}{c}{Group ODf}
        & \multicolumn{2}{c}{Group ED} \\
        & est. & $\sigma$              & est. & $\sigma$     & est. &
$\sigma$          
        & est. & $\sigma$ \\
  \hline
   \hline
$12_0$                   &  -6.4 & 0.3 &   -8.0 & 0.4 &   -6.4 & 0.5 &  
-6.2 & 1.0 \\
$\sigma_{12}$              &   1.7 & 0.1 &   1.2 & 0.2 &   0.6 & 0.4 &  
1.6 & 0.2 \\
$U_0$                   &  -6 & 9 &  -12 & 5 &   -10 & 9 &   -30 & 37 \\
$\sigma_U$              &   22 & 6 &  35 & 8 &  39 & 6 &  106 & 45 \\
$V_0$                   &  -7 & 8 & -26 & 7 &  -24 & 8 &  -97 & 54 \\
$\sigma_V$              &  12 & 5 &  26 & 11 &  22 & 6 &  65 & 64  \\
$W_0$                   &  -9 & 4 & -9 & 5 &  -8 & 5 &  -2 & 44 \\
$\sigma_W$              &  9 & 8 &  21 & 8 &  21 & 7 &  75 & 29  \\
$Z_0$                   &  161 & 55 & 258 & 56 &  256 & 79 &  1065 & 724
\\
$\%$                    &  29 &   & 32 &  &  29 &  &  10 &  \\ 
 \hline
 \end{tabular}
 \end{center}
\end{table}

\begin{table}
\caption{25 calibration: estimated parameters of the different groups
and 
	   percentage of the LPVs population in each of them
        }
\label{tab_esti25}
\begin{center}
\begin{tabular}{|l||rr|rr|rr|rr|}
\hline
        & \multicolumn{2}{c}{Group D} & \multicolumn{2}{c}{Group ODb} 
        & \multicolumn{2}{c}{Group ODf}
        & \multicolumn{2}{c}{Group ED} \\
        & est. & $\sigma$              & est. & $\sigma$     & est. &
$\sigma$          
        & est. & $\sigma$ \\
  \hline
   \hline
$25_0$                   &  -7.1 & 0.5 &   -8.6 & 0.4 &   -6.5 & 0.3 &  
-6.8 & 0.8 \\
$\sigma_{25}$              &   1.7 & 0.1 &   1.2 & 0.2 &   0.6 & 0.4 &  
1.6 & 0.5 \\
$U_0$                   &  -6 & 6 &  -10 & 4 &   -10 & 7 &   -39 & 48 \\
$\sigma_U$              &   21 & 8 &  36 & 10 &  38 & 6 &  111 & 33 \\
$V_0$                   &  -6 & 4 & -26 & 7 &  -22 & 6 &  -99 & 63 \\
$\sigma_V$              &  13 & 4 &  27 & 9 &  22 & 5 &  69 & 23  \\
$W_0$                   &  -10 & 4 & -9 & 5 &  -8 & 4 &  1 & 42 \\
$\sigma_W$              &  11 & 7 &  21 & 8 &  20 & 4 &  75 & 37  \\
$Z_0$                   &  158 & 43 & 277 & 34 &  270 & 107 &  1610 &
1180 \\
$\%$                    &  28 &   & 32 &  &  30 &  &  10 &  \\ 
 \hline
 \end{tabular}
 \end{center}
\end{table}

\subsection{The IRAS bands} \label{Sect:IRAS groups}

The four LPV groups identified in the IRAS 12 and 25 bands are given
in Tables~\ref{tab_esti12} and \ref{tab_esti25}, respectively. They
are similar to those identified in the K band (Table~\ref{tab_estiK}),
except that the old disk group is further divided into ``bright''
(ODb) and ``faint'' (ODf) subgroups. Let us remember here that the 
method  allows us to distinguish groups with 
similar mean kinematics but different luminosities (ODb and ODf)  or
groups with a similar luminosity distribution but different kinematics
(D and ED for instance). Moreover, it is important to remark that D,
ODb and ED have, on average, a similar color index $25-12 = -0.6$ mag
corresponding to a thick circumstellar envelope, while ODf has a mean
index of 0.1 mag that suggests that the majority of the stars in this
last group have thin envelopes.

Let us finally point out that the kinematic parameters
($U_0,V_0,W_0$) associated with each of the four groups are very similar
for both the 12 and 25 calibrations.

\section{Individual estimates and properties of the sample}
\label{Sect:individual}

\begin{figure*}
\centerline{\psfig{figure=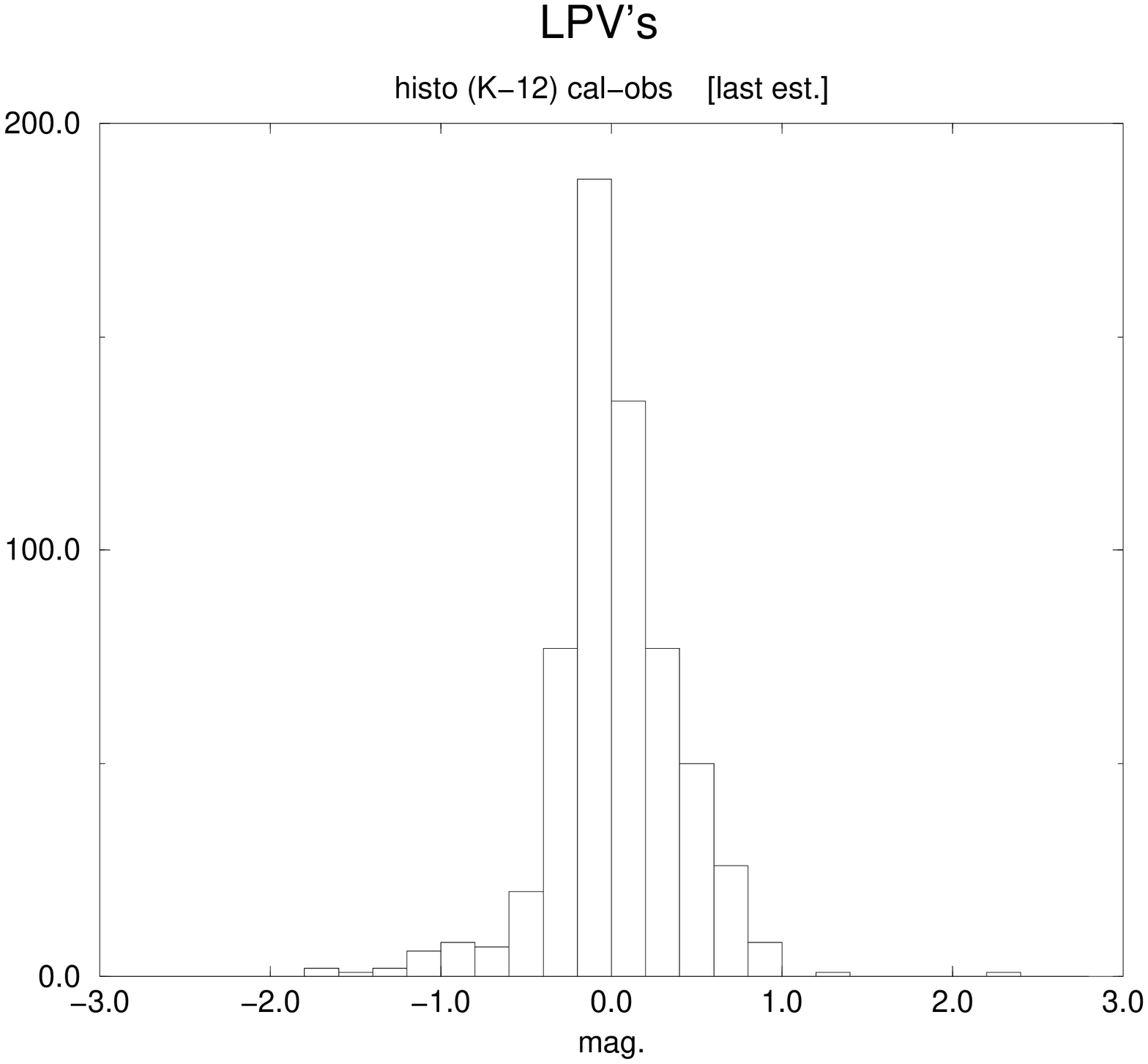 ,height=5cm ,angle=-90}
            \psfig{figure=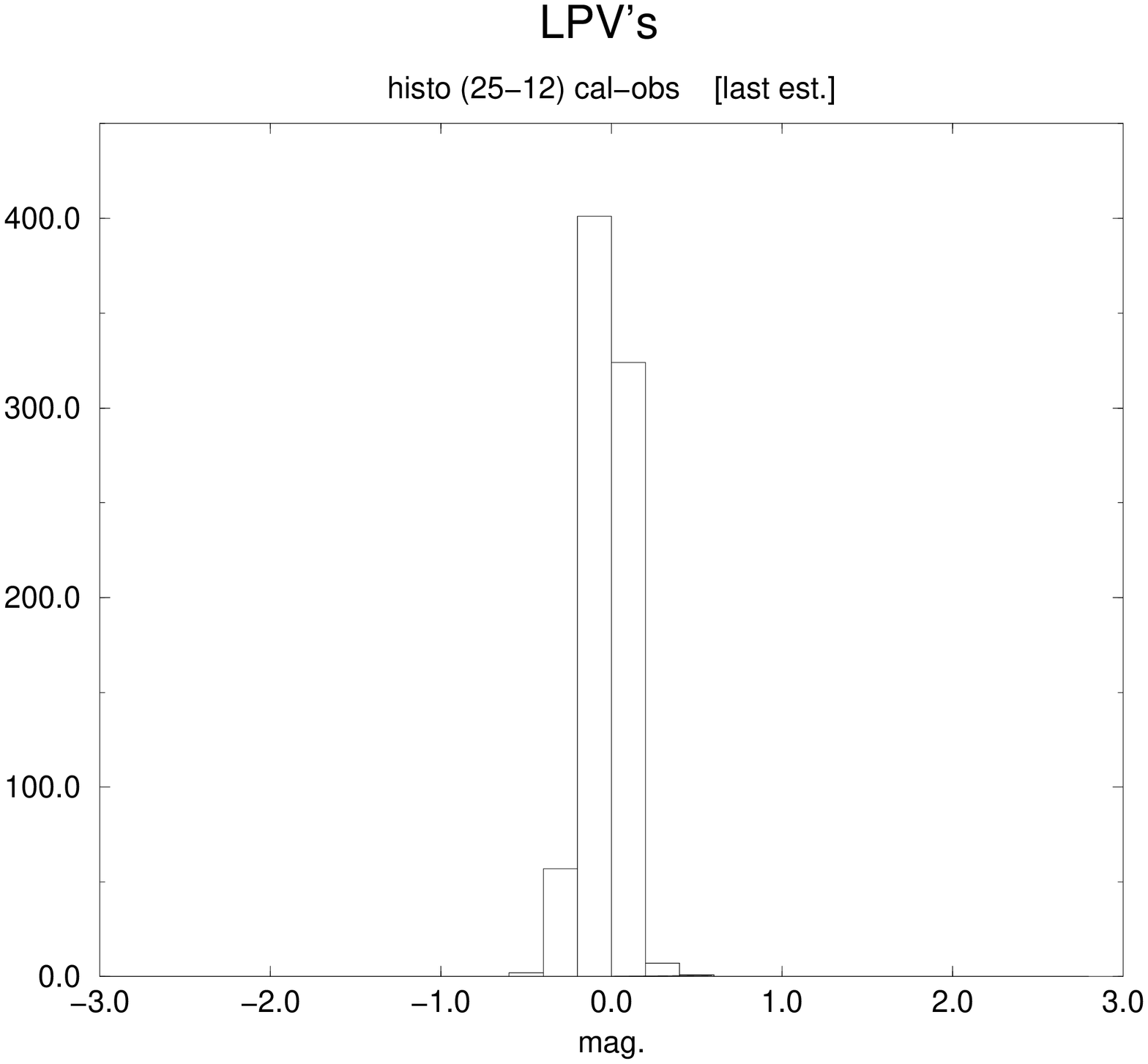 ,height=5cm ,angle=-90}}
\caption{Histograms of the distributions of the differences of the 
         observed color indices ($obs$) and the calculated
         ($cal$) -- arbitrary scale -- from estimated intrinsic
         luminosities for initial and final discriminations of stars
         into the groups 
}
\label{fig_omoinsc}
\end{figure*}

\subsection{Individual estimates} \label{sec_discri}

Once the parameter estimation and group discrimination is completed,
each star in our initial sample is {\it a posteriori} attributed to
one of the LPV groups identified in each bandpass, following the
method described in Sect.~\ref{Sect:statistics}. This allows us to
estimate the most probable individual distance and absolute magnitude
in each band according to the observed astrometric, kinematic and
photometric data and attributed group.

Due to the probabilistic nature of the Bayesian procedure, some
misclassification is unavoidable. To check and improve individual star
assignations in each wavelength, we compare the calculated color
indices $cal=M_{\lambda_1}-M_{\lambda_2}$ (obtained from the estimated
individual absolute magnitudes deduced by the Bayesian assignations in
the $\lambda_1$ and/or $\lambda_2$ wavelengths) with the observed
color indices ($obs=m_{\lambda_1}-m_{\lambda_2}$).  5\% of the stars
are re-assigned to groups reducing the differences $cal-obs$ for their
indices 25-12 and/or K-12.

Figure \ref{fig_omoinsc} shows the histograms of difference of $cal-obs$
for the indices 25-12 and K-12 of all the stars in the sample. These
distributions are related to the errors of the individual estimated
luminosities and of the observed magnitudes.  
 We can deduce that the  accuracies of 
our estimated individual luminosities are distributed 
according to a gaussian rule of standard error
0.3 mag. and 0.1 mag. respectively in K and IRAS bands.
 The lower accuracy in the IRAS bands is
consistent with the fact that IRAS photometry is more homogeneous than
the K photometry, and that the variability amplitude of LPVs is
smaller in the IRAS bands than in K. 
 As previously stated, the LM method has allowed us to take advantage of
all the available information, leading to better estimations of the
individual absolute magnitudes. Furthermore, the LM method has provided
at the same time the statistical distribution of these individual
magnitudes (see Sect. \ref{Sect:statistics} and \ref{Sect:groups}). 
The individual estimates of K, 12
and 25 luminosities are given in a table available in electronic form
at the CDS \footnote {via anonymous ftp to cdsarc.u-strasbg.fr (130.79.128.5)
or via http://cdsweb.u-strasbg.fr/cgi-bin/qcat?J/A+A/} and are included in the specialized
ASTRID database \footnote {via http://astrid.graal.univ-montp2.fr}.

\subsection{Comparison of sample/population} \label{sec_sampop}

The LM method gives unbiased calibrations for the base {\it
population}.  It also gives individual kinematic and photometric
estimates for each star of the {\it sample}.  The distribution of
these individual estimates (Sect.~\ref{sec_discri}) is, of course,
biased by the sample selection criteria, contrary to the group
characteristics derived in Sect.~\ref{Sect:groups}.  A comparison of
the statistical properties of the {\it sample} with the calibrated
parameters for the {\it population} allows us to check the {\it
representativity or the bias of the sample with respect to the
population}.

\subsubsection{Kinematic representativity} \label{sec_kinbias}

Let us analyze the representativity of our sample stars with respect
to the kinematics. The observed proper motions and radial velocities,
together with the estimated individual distances, allow us to compute
the three velocity components $(U,V,W)$ and the distance $Z$ above the
galactic plane for each star in our sample. The mean kinematical
properties of our sample derived from these individual estimates are
shown in Table~\ref{tab_kinbias} for each group of the K and 12 bands.
They are very similar to the parameters describing the groups
(Tables~\ref{tab_estiK} and \ref{tab_esti12}), showing that {\it our
sample is very representative of the LPVs population as far as the
kinematics is concerned}.\\

This conclusion was expected since there is a priori no selection
affecting (directly or indirectly) the kinematics of our sample and
thus no bias is introduced in the kinematics of the stars. We can also
note that the proportion of the different groups in the sample is
close to that found in the population.

\subsubsection{Luminosity representativity} \label{sec_lumbias}

\begin{figure*}

\centerline{\psfig{figure=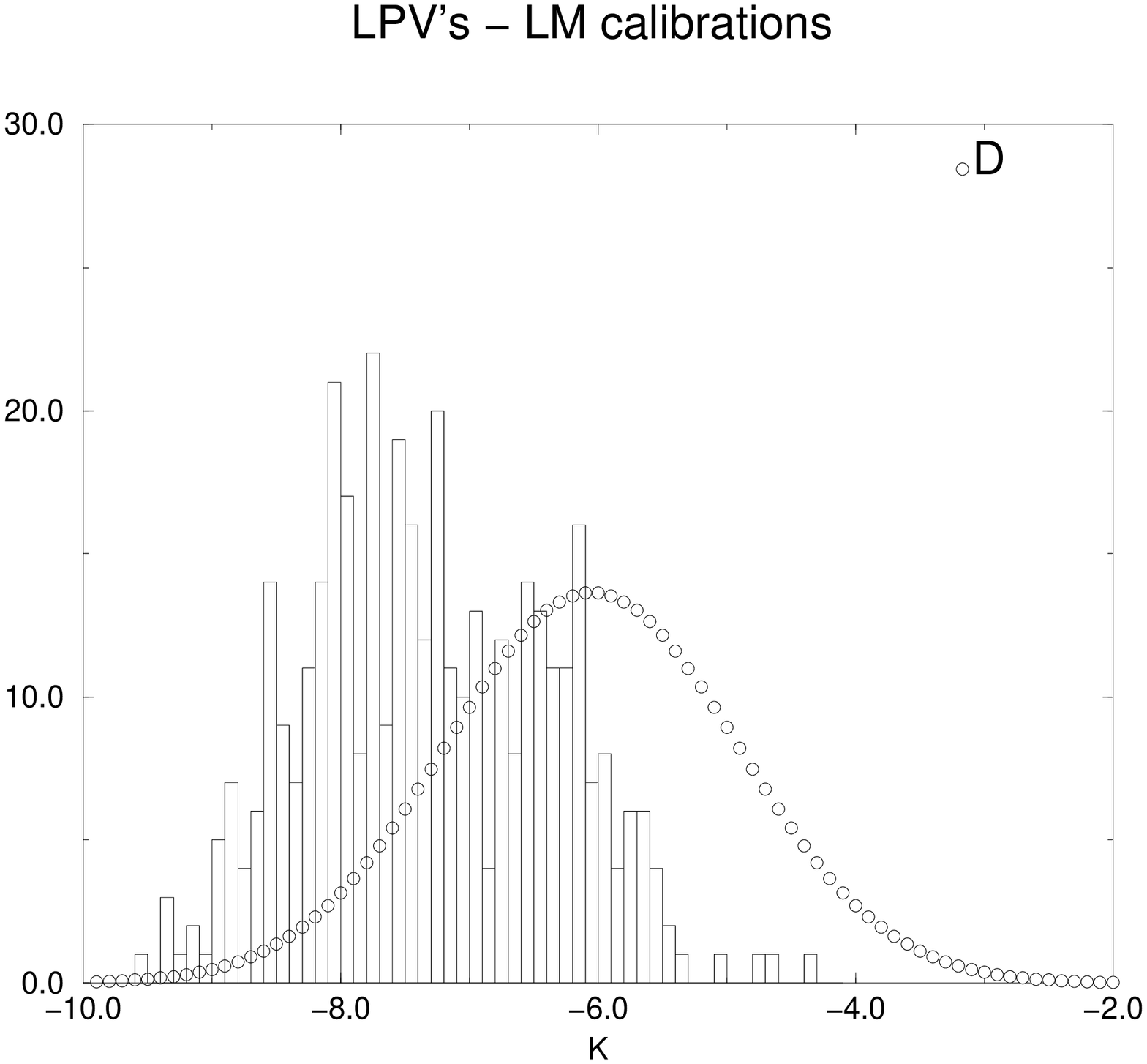,height=4cm ,angle=-90}
            \psfig{figure=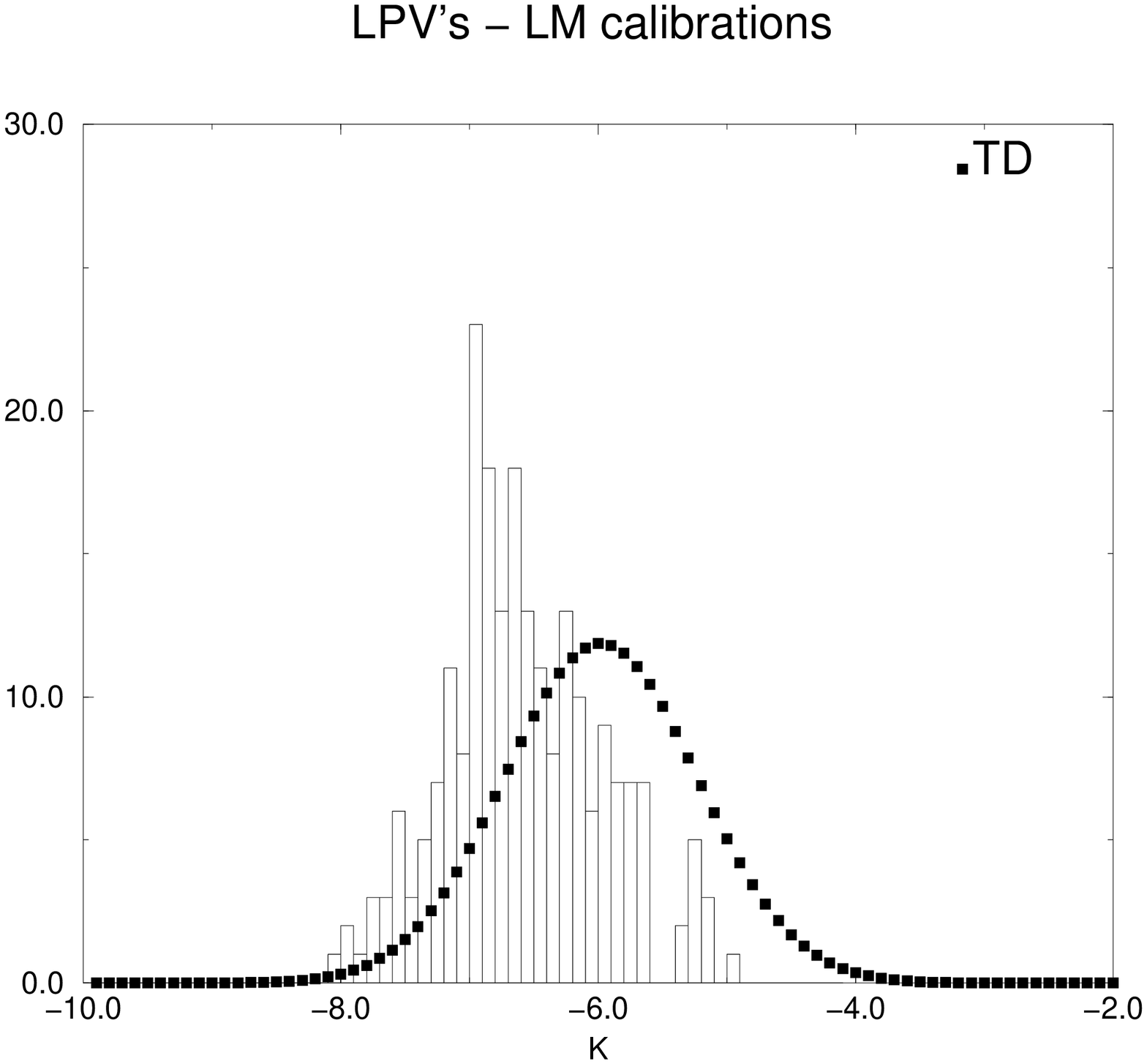,height=4cm ,angle=-90}
            \psfig{figure=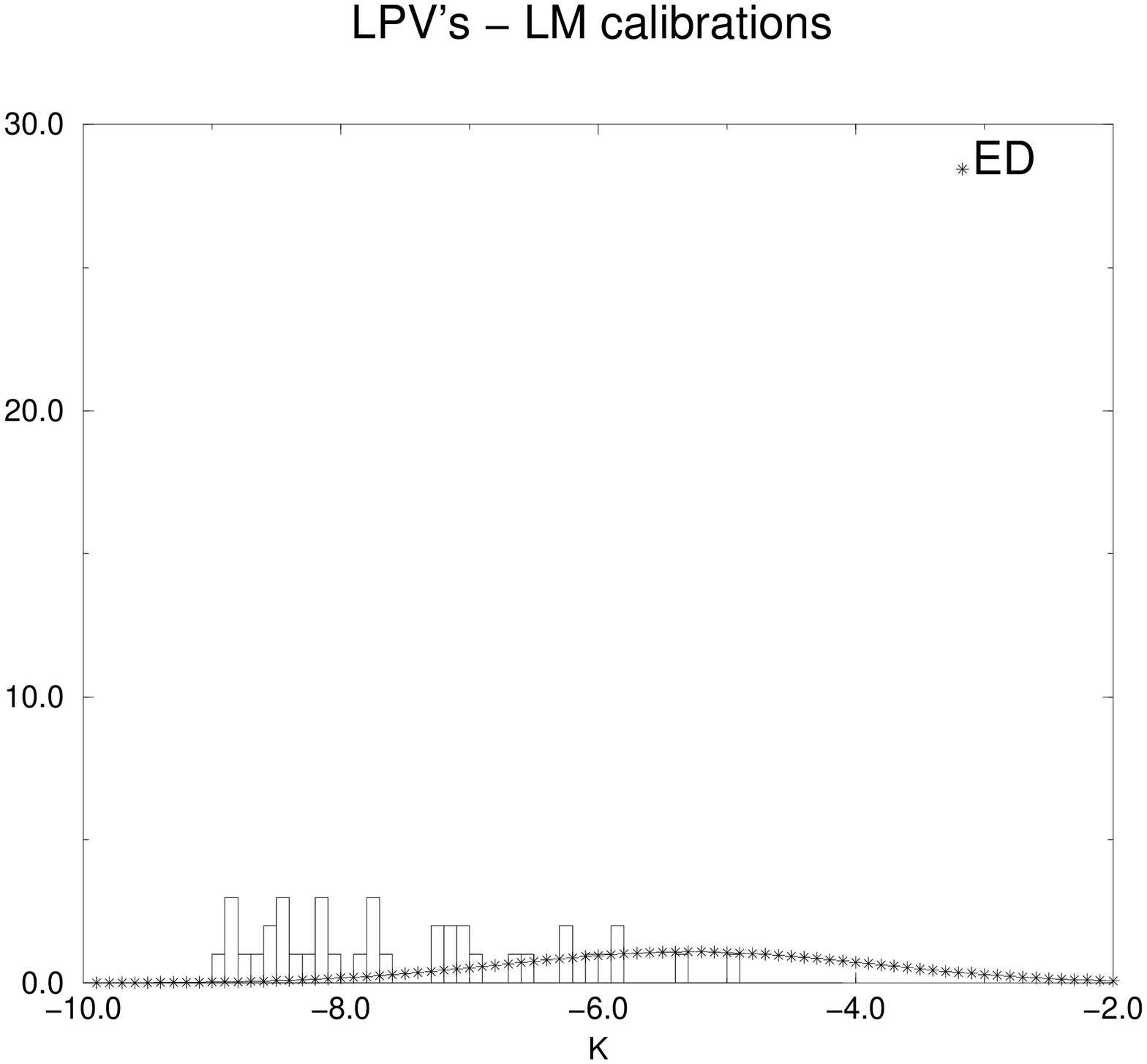,height=4cm ,angle=-90}}
\centerline{\psfig{figure=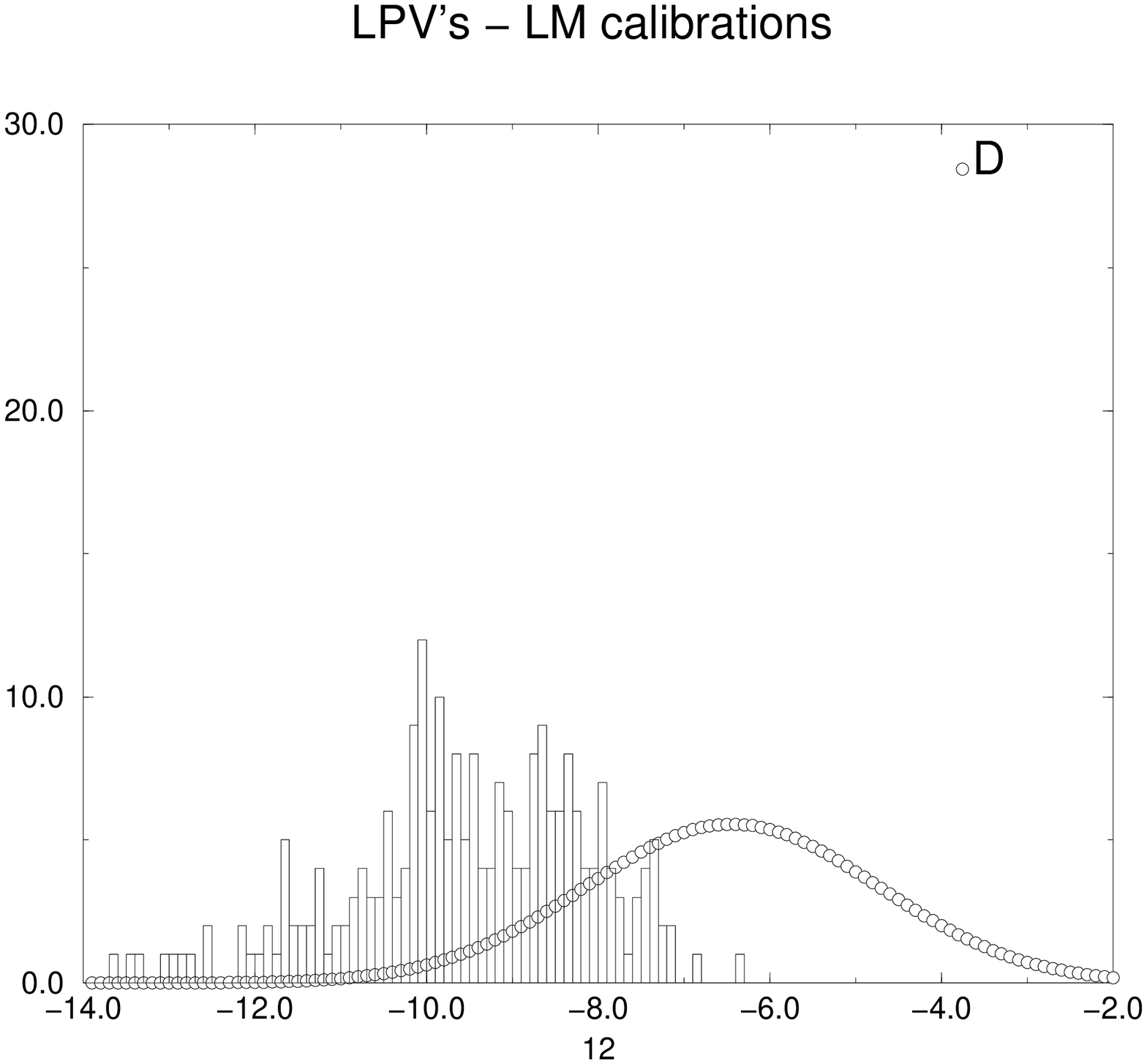,height=4cm ,angle=-90}
            \psfig{figure=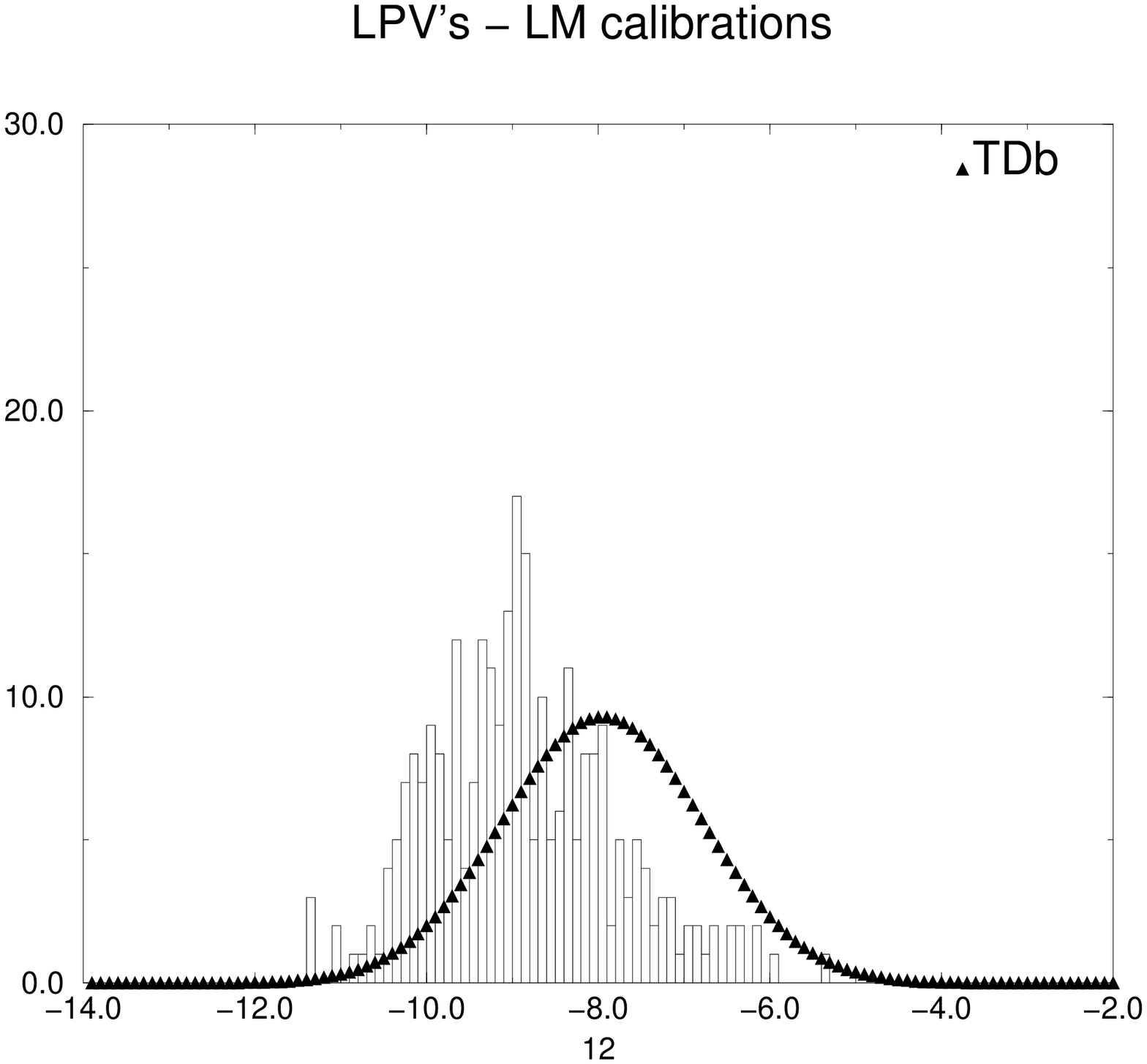,height=4cm ,angle=-90}
            \psfig{figure=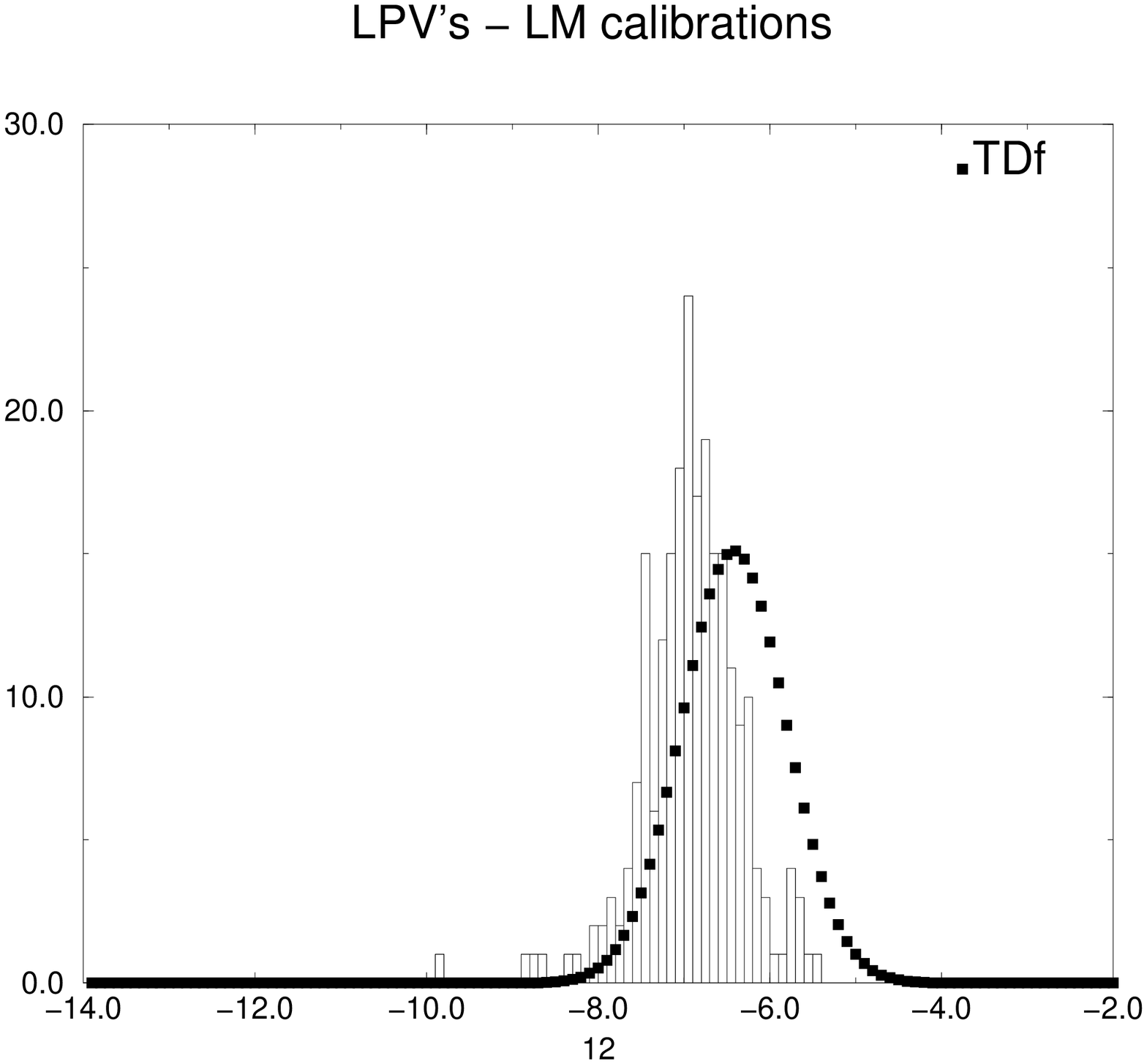,height=4cm ,angle=-90}
            \psfig{figure=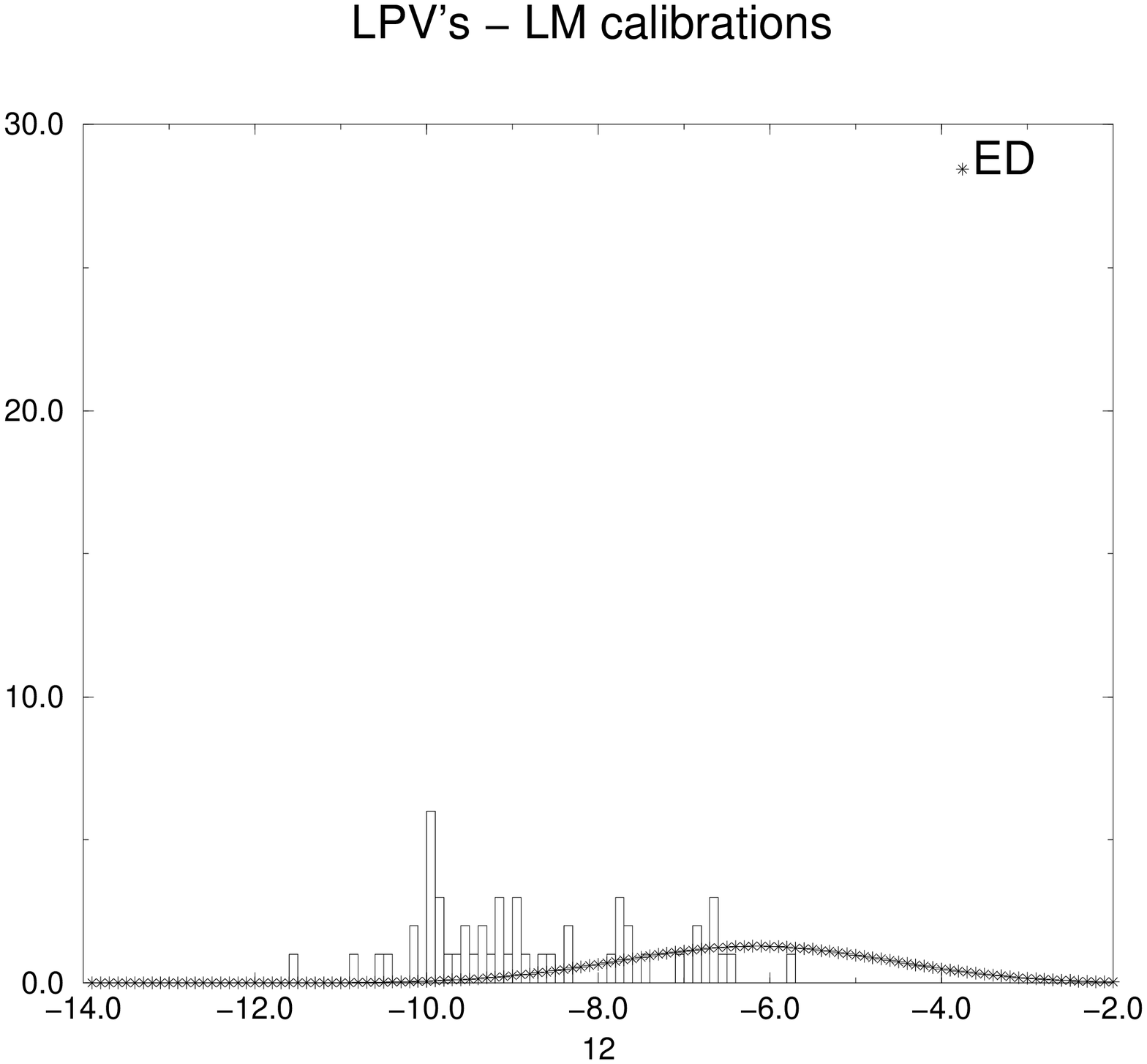,height=4cm ,angle=-90}}
\centerline{\psfig{figure=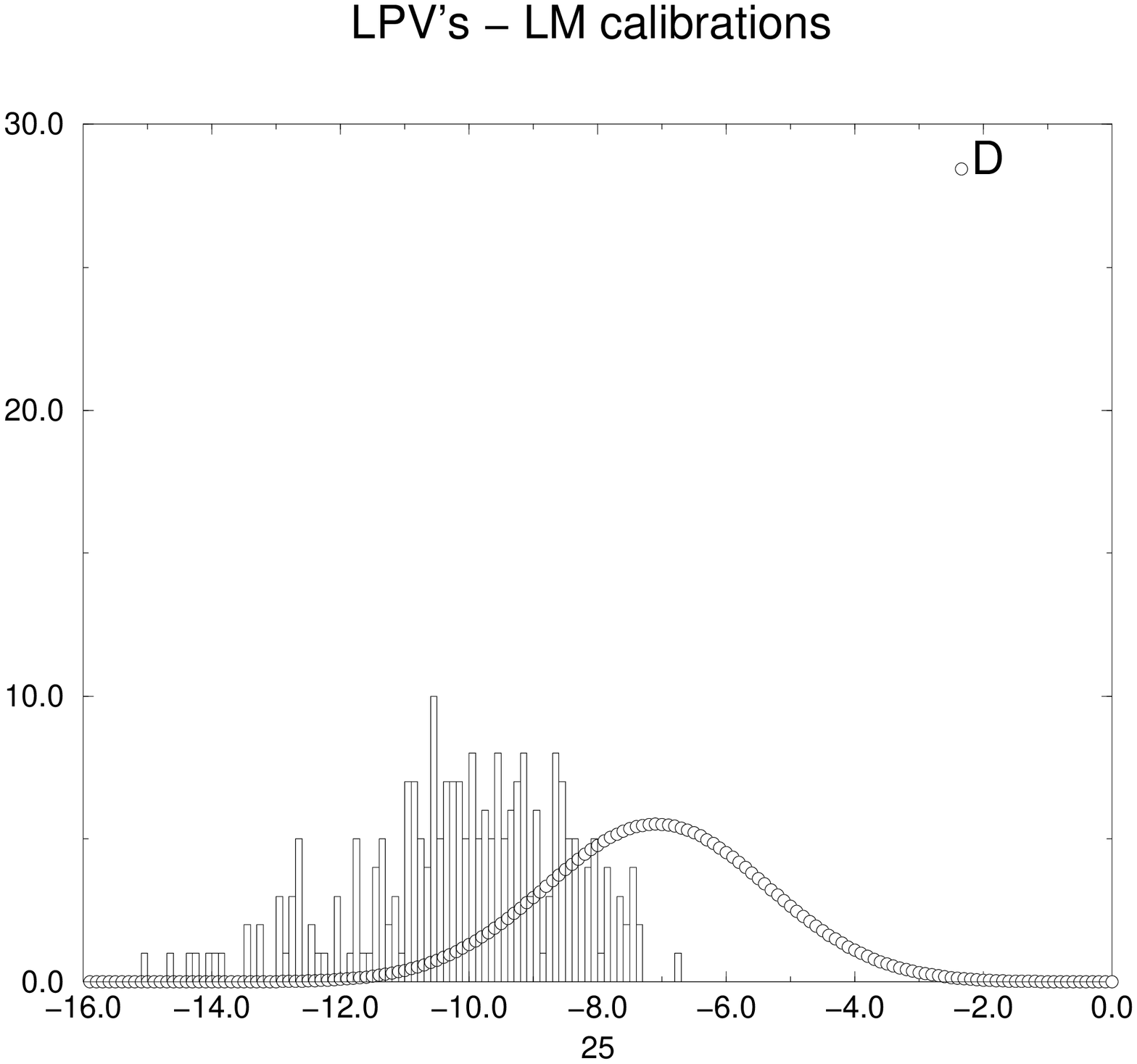,height=4cm ,angle=-90}
            \psfig{figure=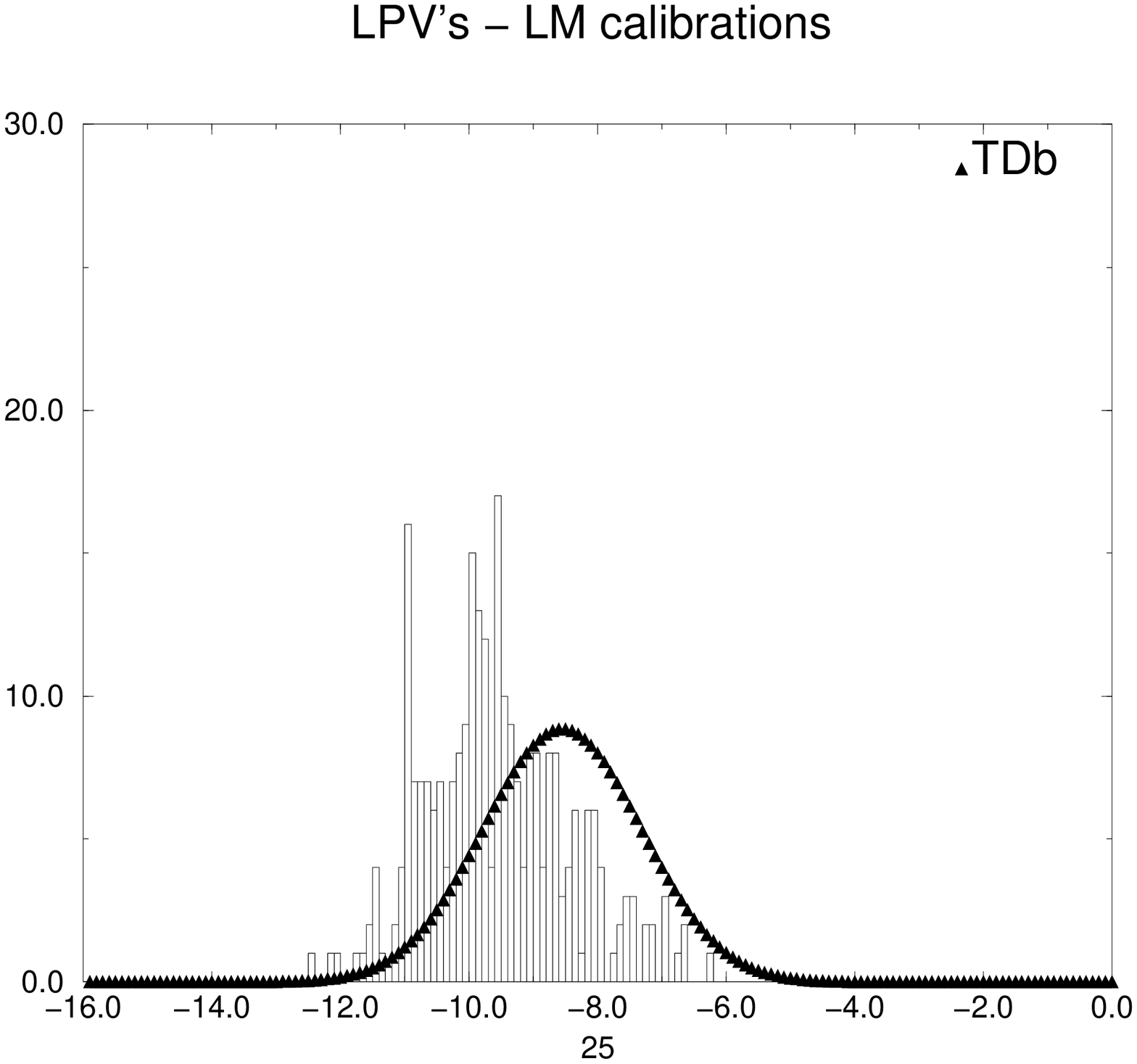,height=4cm ,angle=-90}
            \psfig{figure=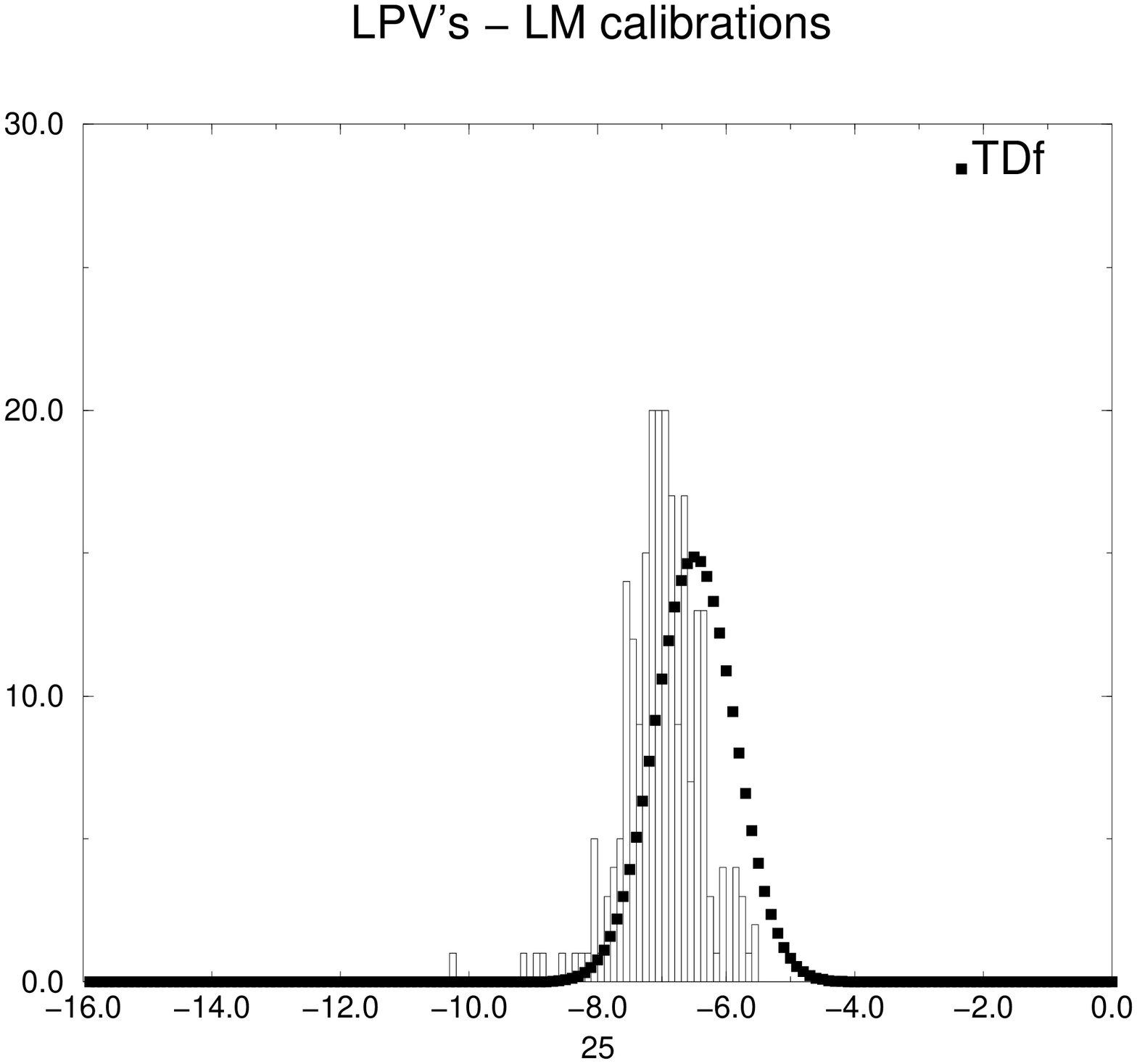,height=4cm ,angle=-90}
            \psfig{figure=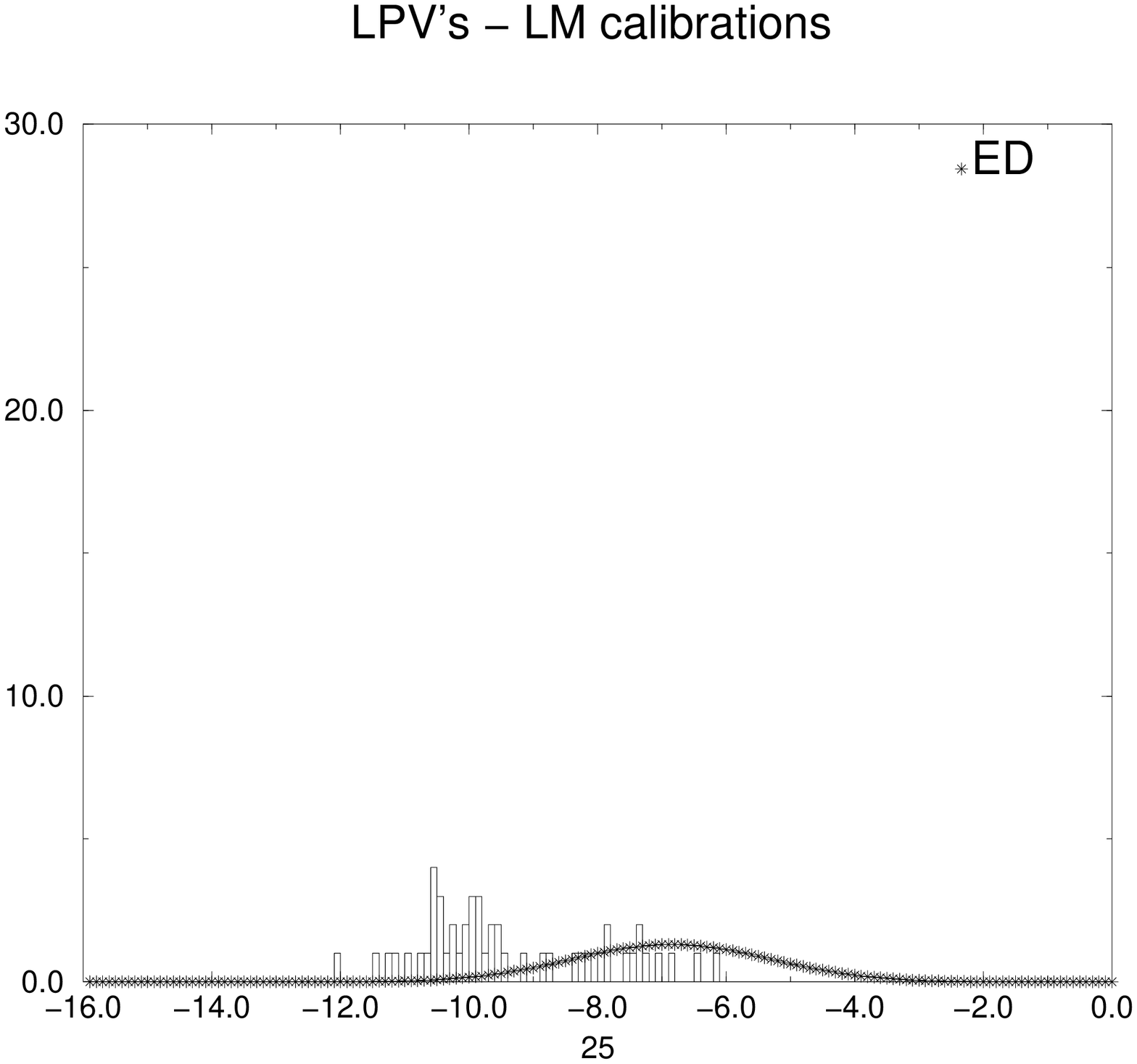,height=4cm ,angle=-90}}
\caption[]{Distributions of individual luminosities in K, 12 and 25 
           from top to bottom, of each group (D,OD ED, or D,ODb,ODf,ED from
           left to right) of the sample compared to the distribution of
           the  calibrated luminosity  (in units normalized to the 
           surface of each histogram) for the same group}
\label{fig_lumbias}
\end{figure*}

Although no kinematical bias is introduced when selecting a sample
with a cut in apparent magnitude, it is well known that a bias in
luminosity is introduced.  Figure~\ref{fig_lumbias} shows the
histograms of the individual absolute magnitudes of the stars in our sample
in each group of the K and IRAS bandpasses, together with the normal
unbiased distributions estimated by the LM method for the population.
 These distributions are in units normalized to the surface of
each histogram -and not to the number of
population stars in each sample -, thus
only the relative shapes and the magnitude shifts of both
histogram and unbiased distribution are relevant. 
The bias of our sample towards higher luminosities is very clear both
in K and IRAS bands. In the IRAS bands, the under-representativity of
faint stars in our sample is more pronounced for LPV stars in the disk
group than in the other IRAS groups. This corresponds to the classical
Malmquist bias (1936), increasing with increased $\sigma_M$
value.  In short, {\it the under-representation of faint stars in our
sample is important for the K or IRAS faint stars and even more for
the disk population, specially in the case of IRAS bandpasses}. \\

However, let us remark that the brightest stars in every group of the
sample coincide with the brightest luminosity of the group base
population.

\begin{table}
\caption{Mean kinematical parameters of the sample computed from
individual 
velocities and positions of stars}
\label{tab_kinbias}
\begin{center}
\begin{tabular}{|l||rrrr|}
   \hline
  & & & & \\
 K & & & & \\
   \hline
    & D & \multicolumn{2}{c}{OD} & ED  \\
  & & & & \\
 $V_0$       & -13 & \multicolumn{2}{c}{-31} & -121 \\
 $\sigma_U$  &  30 & \multicolumn{2}{c}{41}  & 111 \\ 
 $\sigma_V$  &  18 & \multicolumn{2}{c}{27}  & 62 \\
 $\sigma_W$  &  20 & \multicolumn{2}{c}{26}  & 84 \\
 $Z_0$       & 185 & \multicolumn{2}{c}{245} & 621 \\
   N        & 396  & \multicolumn{2}{c}{224}  & 39 \\
   \%        & 60  & \multicolumn{2}{c}{34}  & 6 \\
  \hline
  & & & & \\
 12 & & & & \\
   \hline
    & D & ODb & ODf & ED  \\
  & & & & \\
 $V_0$  & -7 & -28 & -20 & -105 \\
 $\sigma_U$  & 20 & 42 & 35 & 110 \\
 $\sigma_V$  & 14 &  27 & 22 & 66 \\
 $\sigma_W$  & 12 &  39 & 21 & 77 \\
 $Z_0$  & 152 & 343 & 180 & 714 \\
   N   &  239 &  273 &  231 &   51 \\
   \%   &  30 &  34 &  29 &   7 \\
  \hline
  \end{tabular}
  \end{center}
\end{table}

\subsubsection{Envelope effects and representativity}
\label{sec_envbias}

\begin{figure}

\vspace{2cm}
\centerline{\psfig{figure=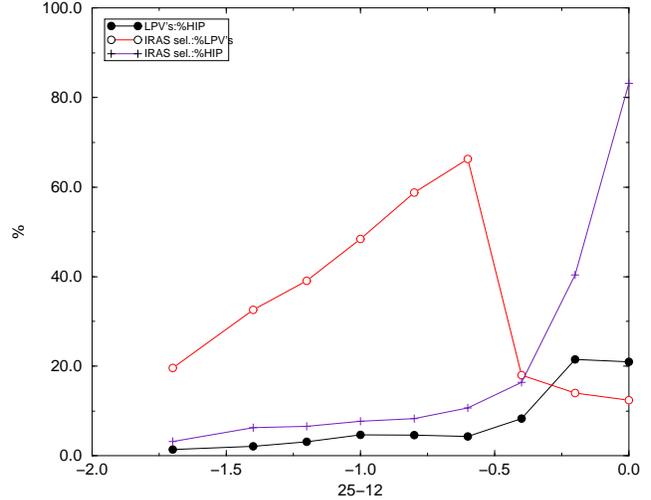 ,height=8cm ,angle=-90}}
\caption{Percentages of LPVs observed by HIPPARCOS (full circles) 
as a function of the 25-12 IRAS color index. They are compared to 
the percentages of stars observed by HIPPARCOS (+) and 
of known LPVs (*)
among the sample of IRAS sources selected as probable LPVs from their 
IRAS color indices. }
\label{fig_irasgcvs}
\end{figure}

The luminosity sampling bias is not independent of the existence,
thickness and composition of a circumstellar envelope around LPVs.
Figure \ref{fig_irasgcvs}, which shows the percentage of known LPVs
measured by HIPPARCOS (LPVs:\%HIP)  as a function of the IRAS (25-12)
color index, shows that the incompleteness depends on the
IRAS color. This is not surprising because the thicker the envelope,
the fainter the star in the visual wavelengths. 

This is confirmed if instead of using the known LPVs we use the IRAS
sources with a (25-12) color index compatible with the LPVs values of
this index. In doing so, we include stars in the LPV region not
necessarily classified as variables (IRAS sel.:\%LPVs). The bias of
the HIPPARCOS sample is more strongly dependent on the envelope
thickness if we do the comparison with these selected IRAS sources.
Thus the percentage of stars observed by HIPPARCOS (IRAS sel.:\%HIP)
strongly and abruptly increases up to 80 \% for 25-12 decreasing to
zero.\\

\begin{figure}
\vspace{2cm}
\centerline{\psfig{figure=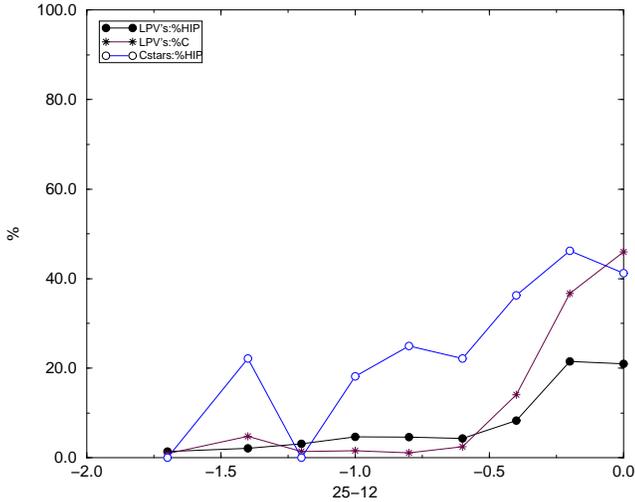 ,height=8cm ,angle=-90}}
\caption{Percentages of C stars observed by HIPPARCOS (empty circles) 
and among known LPVs (empty squares) 
as a function of the 25-12 IRAS color index compared with the 
percentages of LPVs observed by HIPPARCOS (full circles). }
\label{fig_Cbias}
\end{figure}

Finally, figure \ref{fig_Cbias} shows how much the sample
of carbon-rich stars measured by HIPPARCOS (C stars:\%HIP) does not
represents either the percentage of the C-rich stars among the known
LPVs (LPVs:\%C stars)  or the percentage of stars known as LPVs
measured by HIPPARCOS (LPVs:\%HIP).  Thus one should be careful about
making any interpretation from the percentages of C-rich stars, as we
will
see in sect. \ref{sec_vartype} .

\section{Properties of crossed groups} \label{sec_propgrups}

\subsection{Crossing K and IRAS} \label{Sect:crossing}

Due to the difficulties coming from the large uncertainties on $m_V$:
large amplitudes and variations from one cycle to another, the V
results will not be further used in our analysis and we will
concentrate on the K and IRAS results.\\

The luminosity estimations in the K and IRAS bands complement each
other in the sense that, in general, K fluxes characterize stellar
properties while IRAS fluxes provide information on the circumstellar
envelope.  Thus, the most physically interesting results are obviously
obtained by simultaneously considering K and IRAS luminosities. We
have already seen that it is very difficult to calibrate these
luminosities at the same time due to the incomplete knowledge and
non-uniqueness of the relation between the different magnitudes
(Sect.~\ref{Sect:statistics}).  Another possibility is to do a
crossing of the groups from both the K and IRAS calibrations i.e. to
examine the properties of the stars belonging to the same group in K
and IRAS.

The first remarkable result concerns the number of crossed groups:  
only 7 are not empty while 12 could, a priori, be expected.
Interestingly, there is no mixture of the extended disk group (in
either wavelength) with any other group, except two stars (O-rich SRa:
RW Eri and O-rich Mira: SV And), which is compatible with the
statistical classification errors. This is a nice confirmation of the
power of the LM method to extract consistently distinct groups in
biased samples of a given stellar population.

\begin{table*}
\caption{Mean kinematical parameters for the crossing K(IRAS) groups 
computed from individual velocities 
 and positions. They can be considered as representative for the LPVs 
population (see Sect.~\ref{Sect:crossing} }
\label{tab_kincross}
\begin{center}
\begin{tabular}{|l||rrrrrrr|}
   \hline
  & & & & & & &  \\
  & \multicolumn{2}{c}{Disk1} & \multicolumn{2}{c}{Disk2} &
\multicolumn{2}{c}{Old Disk} & Ext. Disk\\
  & & & & & & &  \\
  &  D(D) & OD(D) & D(ODf) & D(ODb) & OD(ODf) & OD(ODb) & ED(ED)\\
   \hline
   \hline
 nb of stars &  141 & 21 & 103 & 90 & 81 & 113 & 36 \\ 
 $V_0$  &  -6 & -7 & -18 & -19 & -34 & -32 & -123 \\ 
 $\sigma_U$  &  23 & 28 & 35 & 36 & 42 & 40 & 114 \\ 
 $\sigma_V$  &  13 & 23 & 20 & 18 & 24 & 26 & 63 \\ 
 $\sigma_W$  &  11 & 11 & 34 & 16 & 21 & 29 & 80 \\ 
 $Z_0$  &  166 & 191 & 208 & 231 & 160 & 310 & 620 \\ 
   \hline
   \hline
  & & & & & & &  \\
 age range & \multicolumn{2}{c}{1-4 $10^9$ yr} & \multicolumn{2}{c}{4-8
$10^9$ yr} & \multicolumn{2}{c}{8-gt10 $10^9$ yr} & \\
  & & & & & & &  \\
 lower mass limit & \multicolumn{2}{c}{2-1.4 ${\cal M}_{\sun}$} &
\multicolumn{2}{c}{1.4-1.15 ${\cal M}_{\sun}$} &
\multicolumn{2}{c}{1.15-lt1 ${\cal M}_{\sun}$} & \\
   \hline
  \end{tabular}
  \end{center}
\end{table*}

Table~\ref{tab_kincross} gives the number of stars in our sample which
are assigned to every crossed group G(G') i.e. to group G in K and G'
in IRAS.  In Sect.\ref{sec_kinbias} our LPVs sample is shown to be
representative of the LPVs population as far as the kinematics is
concerned. Thus {\it the mean kinematics of the stars belonging to a
crossing group K(IRAS) may be considered as representative of the mean
kinematics of the LPVs population belonging to this group}. \\

Obviously such a consideration does not apply to the luminosities (see
Sect.\ref{sec_lumbias}). The assigned groups are given in annex A
(electronic table).

\subsection{Ages and initial masses} \label{sec_age}

Table \ref{tab_kincross} gives the values of the axes of the velocity
ellipsoids and the scale height of each of the 7 crossed K and IRAS
groups.  Given that our sample is representative of the population in
terms of kinematics, as already seen in Sect. \ref{sec_kinbias}, we
can use the kinematical values of table \ref{tab_kincross} as
representative in terms of galactic populations.\\

The relation between the mean kinematics of a galactic population and
its age allows us to estimate the range of ages of the groups.  
Furthermore, classical statistical studies of stars known to belong to
different galactic populations and of different metallicity abundances
allow us to add an estimate of the range of metallicity. By comparing
the values in table \ref{tab_kincross} with the results on kinematics
and metallicity of the galactic populations by Mihalas and Binney
(1981) and by Stromgren (1987), we can deduce:

\begin{itemize}

\item  D(D) and OD(D) populations are 1-4 $10^9$ yr old with a solar
       metallicity. Part of the stars in D(D) are classified as
belonging
       to the bright disk population (BD) by the V analysis, being 
       younger and with a small over-abundance.

\item  The age of D(ODf) and D(ODb) populations can be estimated 
       to be in the range 4-8 $10^9$ yr, with a metallicity from the 
       solar one to [Fe/H] around -0.4 i.e. Z between 0.006 and 0.02. 

\item  OD(ODf) and OD(ODb) populations are composed of stars older
       than 8 $10^9$ yr, up to more than $10^{10}$ yr, 
       with [Fe/H] from the solar value to -0.7 i.e. Z between 0.004
       and 0.02
 
\item  Stars classified as belonging to ED are probably very old and
       deficient with [Fe/H] between -0.7 and -1.5 i.e. Z of the
       order of (0.001,0.004).

\end{itemize}

Moreover, we can estimate initial masses from evolutionary tracks. From
Binney and Merrfield (1998) we can estimate a lower limit of the
initial mass of a given age star that has reached the AGB. Thus, values
of 2, 1.4, 1.15, 1 ${\cal M}_{\sun}$ can be deduced as lower limits of
${\cal M}_{ms}$ of the stars of solar metallicity of respectively 1,
4, 8, 12 $10^9$ yr. This agrees with the results on the ages at the
top of the early-AGB (Charbonnel et al.,1996).

All these results are in the same ranges as the ones given by Jura and
Kleinmann (1992), but our classification is more refined because it is
not based on the spectral types and periods which are now known to not
really be discriminative parameters for LPVs.

In the rest of this paper the stars belonging to D(D) and OD(D),
D(ODf) and D(ODb), OD(ODf) and OD(ODb), ED(ED) will be called disk1,
disk2, old disk and extended disk LPVs respectively (see tables
\ref{tab_kincross} and \ref{tab_repul}).

\subsection{Evolutionary tracks } \label{sec_evtracks}

\begin{figure*}

\centerline{\psfig{figure=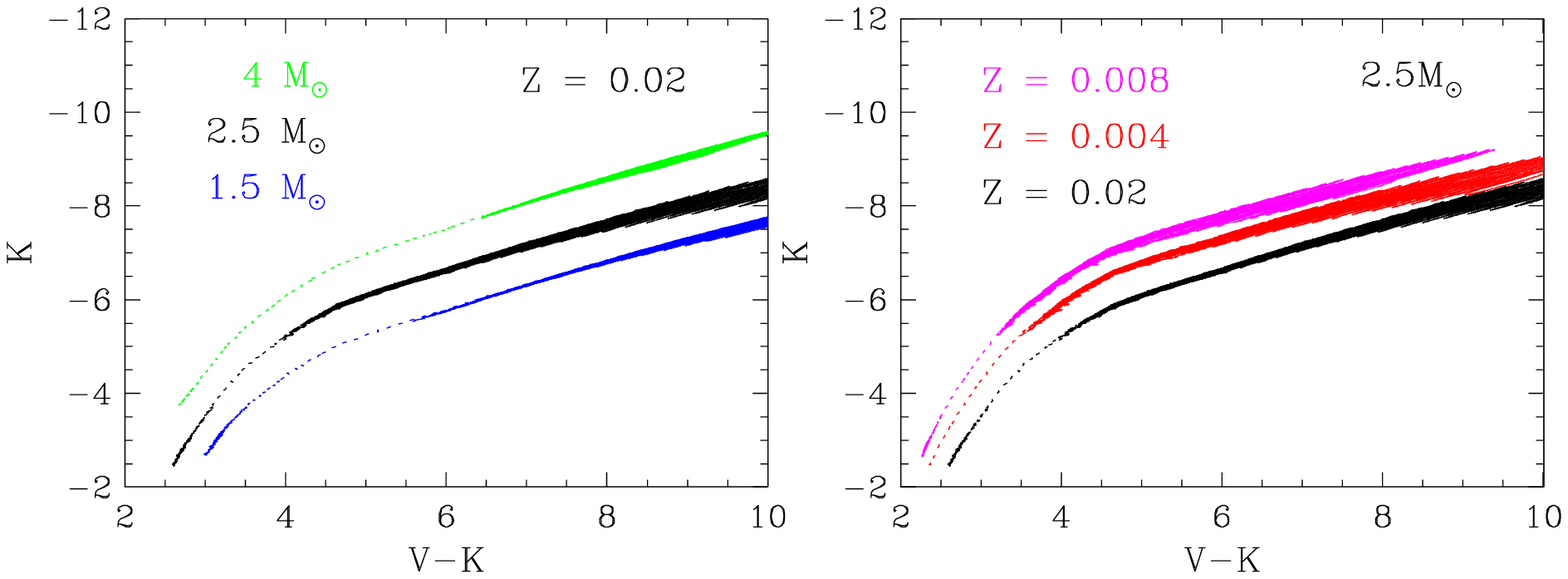 ,height=6.5cm}}

\centerline{\psfig{figure=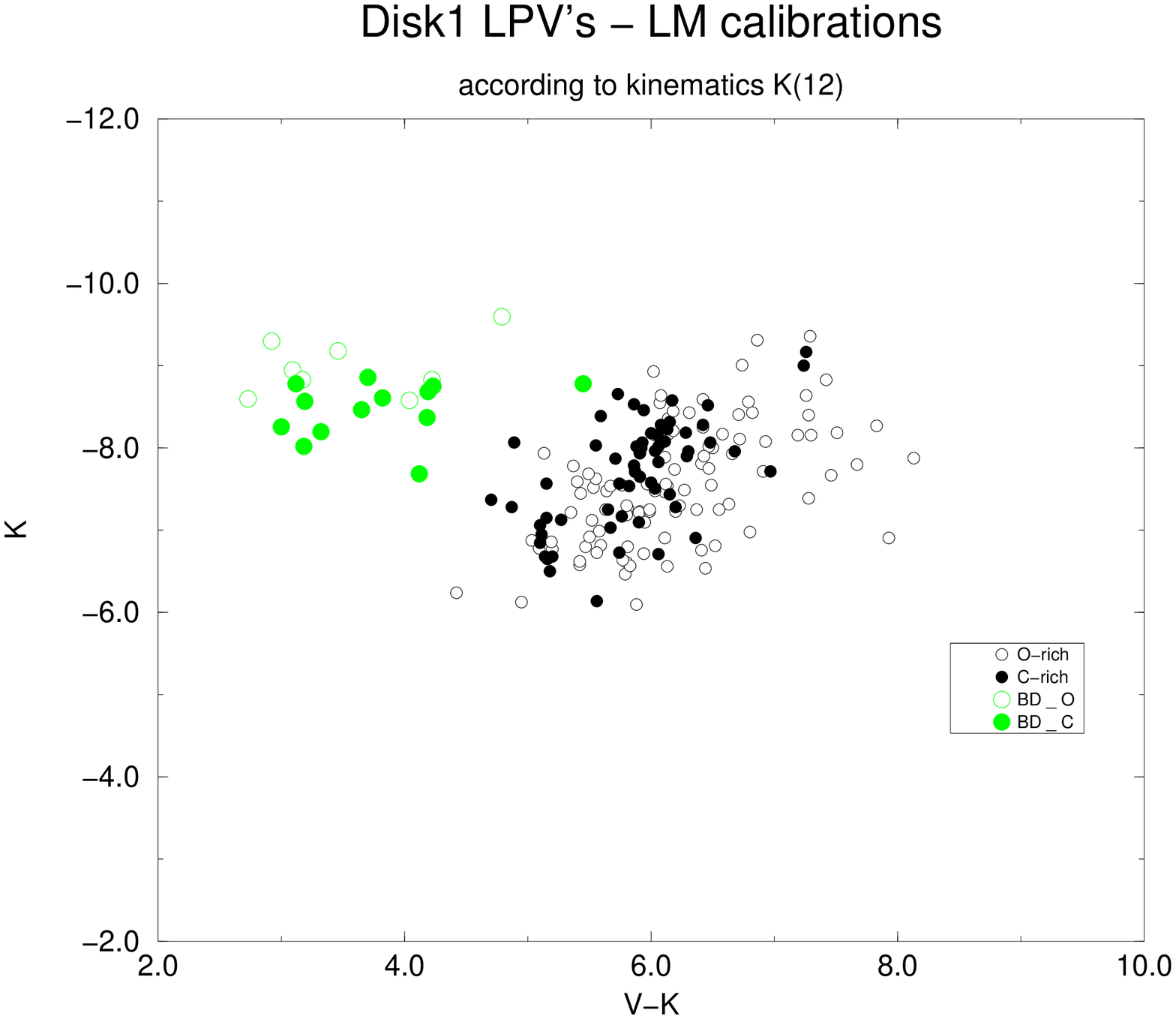,height=8cm ,angle=-90}
            \psfig{figure=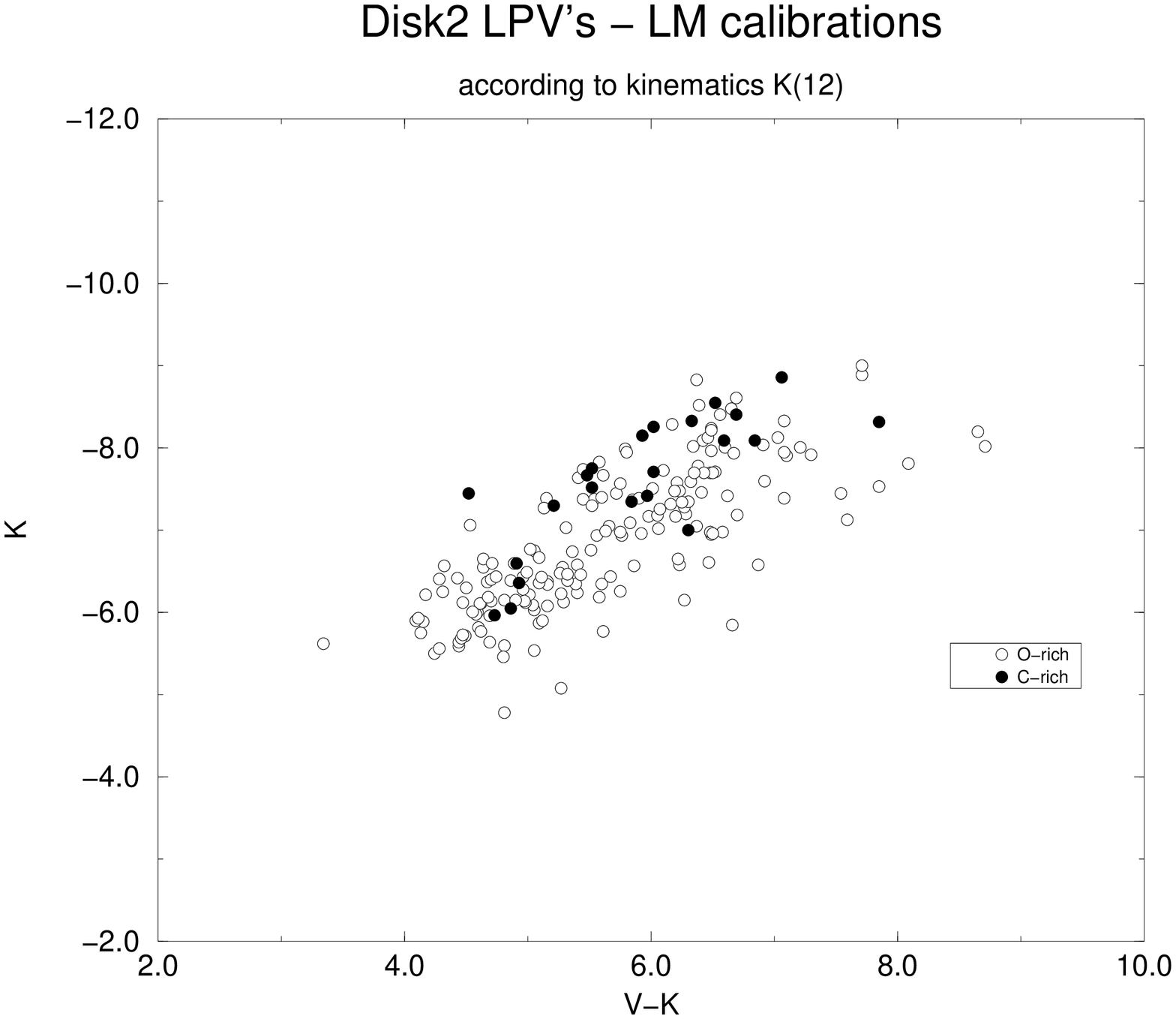,height=8cm ,angle=-90}}

\centerline{\psfig{figure=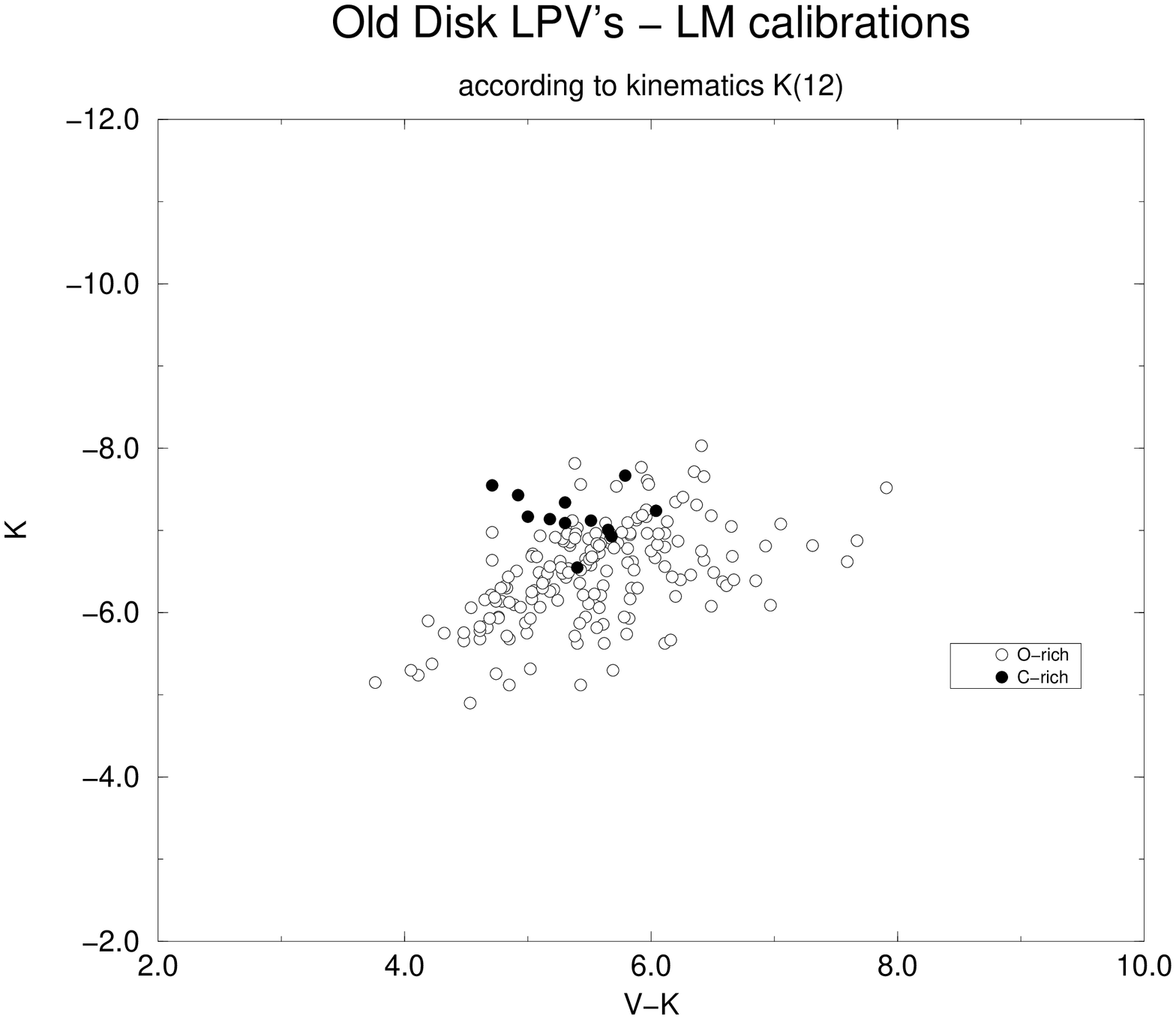,height=8cm ,angle=-90}
            \psfig{figure=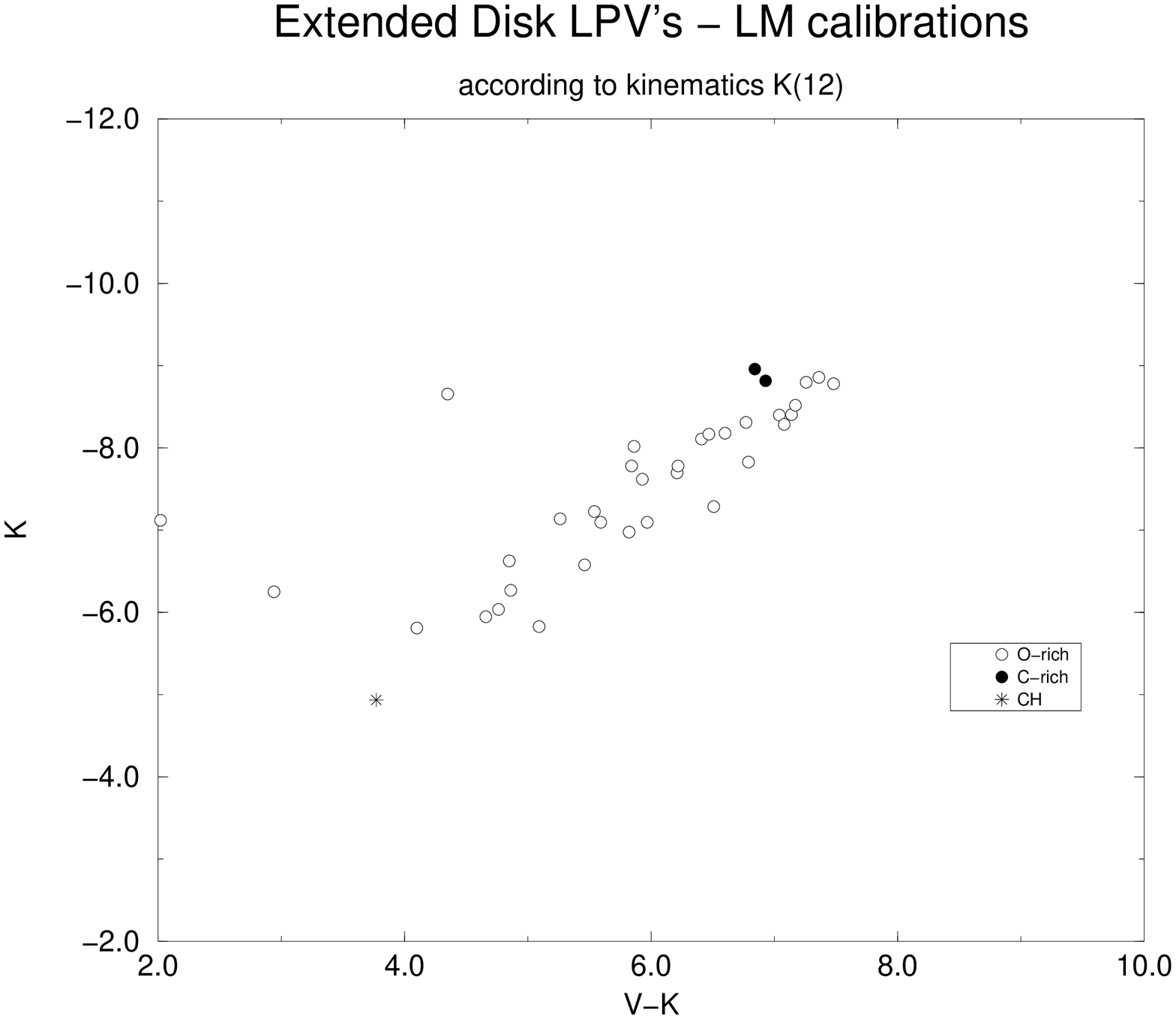,height=8cm ,angle=-90}}

\caption{Theoretical evolutionary AGB tracks for stars of 1.5, 2.5 and 4
         ${\cal M}_{\sun}$ with solar metallicity and for deficient
         (Z=0.008 and Z=0.004) stars of 2.5 ${\cal M}_{\sun}$ compared
         to distribution of the individual estimated K luminosities as a 
         function of the color index V-K according to the assigned
         kinematical groups.}

\label{fig_trackV-K}

\end{figure*}

Figure~\ref{fig_trackV-K} shows the K magnitude as a function of the
V-K color index for each of the disk1, disk2, old disk and extended
disk groups. For comparison, evolutionary AGB model predictions are
also shown for three different masses (1.5, 2.5 and \mass{4}) at solar
metallicity, and at three different metallicities (Z=0.004, 0.008 and
0.02) for \mass{2.5} stars. These models have been computed at Geneva,
and are described in Mowlavi (1999) and Mowlavi \& Meynet (2000). The
conversion between model variables (effective temperature $T_{eff}$
and luminosity $L$) to observable quantities ($V-K$ and $M_K$) was
done by using the transformations given by Ridgway et al. (1980).

Several uncertainties affect both model predictions and the color
transformation relations for AGB stars (which are characterized by
peculiar chemical compositions as a result of dredge-up episodes).
They also affect the determination of the stellar V magnitudes as
noted in Sect.~\ref{Sect:V groups}. Thus, the comparison between
the evolutionary tracks and the distributions of our sample stars in
each group shown in Fig.~\ref{fig_trackV-K} can only provide
qualitative results.

Evolutionary tracks show that, at a given metallicity, stellar
luminosities increase with initial stellar mass (at a given V-K). We
thus conclude, at least qualitatively and due to the solar or slightly
deficient abundance of the majority of HIPPARCOS stars, that our
sample stars have lower mass limits which respectively decrease as we
consider disk1, disk2 and old disk groups. This is in agreement with
the conclusions drawn in Sect.~\ref{sec_age}. Stars of the extended
disk group, on the other hand, are compatible with lower metallicities, 
given their higher K luminosities.

We note that the evolutionary tracks cannot, even if freed from any
uncertainty, attribute a single (M,Z) set of parameters to a star
because of the degeneracy of those two parameters. A higher luminosity
in K at a given V-K could be either attributed to a higher initial mass
or lower metallicity. Kinematics can help in distinguishing such
ambiguous cases.

Finally, we can comment on the smaller V-K values for the bright disk
LPVs.  This confirms the strong circumstellar absorption in V for
these massive stars.

\subsection{Variability and spectral types} \label{sec_vartype}

The composition of each crossed group K(IRAS) with respect to usual
classifications of LPVs (variability and spectral types) is given by
the contingency table of both assignations (Table
\ref{tab_contengency}). The associated attraction-repulsion indices
(Tenenhaus, 1994)  -- ratio of the observed frequency to the
theoretical frequency in the case of independence of both modalities
-- are more significant in characterizing the correspondence analysis of
types and of groups. These indices are given in Table
attract or repel each other if the attraction-repulsion index is
larger or smaller than 1 respectively.

\begin{table*}

\caption{Contingency table between crossing groups and 
L, SR and M variability and M and C spectral types }

\label{tab_contengency}

\begin{center}
\begin{tabular}{|l||rrrr|rrrr|}
   \hline
    & L-C & SRb-C & SRa-C & M-C & L-O & SRb-O & SRa-O & M-O \\
   \hline
   \hline
  D(D) & 29 & 25 & 7 & 9 & 20 & 29 & 4 & 31 \\
  OD(D) & 1 & 3 & 0 & 1 & 1 & 4 & 1 & 9 \\
  D(ODb) & 7 & 3 & 1 & 4 & 6 & 33 & 7 & 28 \\
  D(ODf) & 2 & 4 & 0 & 2 & 42 & 39 & 3 & 2 \\
  OD(ODf) & 3 & 5 & 0 & 1 & 34 & 22 & 2 & 0 \\
  OD(ODb) & 0 & 3 & 0 & 0 & 14 & 28 & 13 & 54 \\
  ED(ED) & 1 & 1 & 0 & 1 & 4 & 7 & 3 & 18 \\
   \hline
  \end{tabular}
  \end{center}
\end{table*}

\begin{table*}

\caption{Attraction-repulsion indices between crossing groups and 
L, SR and M variability and M and C spectral types}

\label{tab_repul}

\begin{center}
\begin{tabular}{|l|l||rrrr|rrrr|}
   \hline
  &  & L-C & SRb-C & SRa-C & M-C & L-O & SRb-O & SRa-O & M-O \\
   \hline
   \hline
Disk1 &  D(D) & {\bf 2.7} & {\bf 2.3} & {\bf 3.5} & {\bf 2.0} & {\it
0.6} & 0.7 & {\it 0.4} & 0.9 \\
Disk1 &  OD(D) & 0.7 & {\bf 1.4} & {\it 0} & {\bf 1.6} & {\it 0} & 0.9 &
0.9 & {\bf 1.9} \\
Disk2 &  D(ODb) & {\it 0.3} & {\it 0.5} & 0.8 & {\bf 1.5} & {\it 0.3} &
{\bf 1.4} & {\bf 1.4} & {\bf 1.4} \\
Disk2 &  D(ODf) & {\it 0.3} & {\it 0.6} & {\it 0} & 0.7 & {\bf 2.3} &
{\bf 1.5} & {\it 0.6} & {\it 0.1} \\
Old Disk &  OD(ODf) & {\it 0.6} & 1.1 & {\it 0} & {\it 0.5} & {\bf 2.5}
& 1.2 & {\it 0.4} & {\it 0} \\
Old Disk &  OD(ODb) & {\it 0} & {\it 0.4} & {\it 0} & {\it 0} & {\it
0.6} & 0.9 & {\bf 2.2} & {\bf 2.1} \\
Ext. Disk &  ED(ED) & {\it 0.4} & {\it 0.4} & {\it 0} & 0.9 & {\it 0.6}
& 0.7 & 1.6 & {\bf 2.2} \\
   \hline
  \end{tabular}
  \end{center}
\end{table*}

>From Table ~\ref{tab_repul} we can deduce that:

\begin{itemize}

\item The groups corresponding to initially less massive stars are
      essentially attractive for O-rich LPVs and the ones
      corresponding to initially massive stars are essentially
      attractive for C-rich LPVs.

\item The strong attraction of C-rich stars by groups
      corresponding to more massive initial stars is clear. It agrees
      with the results already given by V calibration (Mennessier et
      al., 1999)  and with both observations and model predictions 
      according to which dredge-up of carbon from core to the 
      surface of AGB stars is more efficient in massive stars 
      (Dopita et al.,1997). However, we must
      be very cautions. Indeed, the biases described
      before (section \ref{sec_envbias})  show the 
      over-representativity of C-rich stars in the sample and the
      large number of missing LPVs. To which galactic population do
      the not-HIPPARCOS-observed LPVs belong?  Are they O or
      C-rich? The ratio of carbon to oxygen-rich LPVs that goes up
      to 78\% for the D(D) sample could be due to the sampling.  Thus
      deduced implications for the dependence on the mass of the
      efficiency of the dredge-up are to be taken with caution.

\item O-rich SRa's are distributed close to O-rich Miras.

\item O-rich SRb's are significantly present in the two disk
      population groups. The first one is composed of stars without
      a shell or with a thin one (cf sect.\ref{Sect:IRAS groups}). This
      probably corresponds to early AGB stars with a relatively high
      initial mass. The second one contains stars with a
      shell and corresponds to stars in the same stage as SRa and
      Miras. This agrees with results by Kerschbaum and Hron
      (1992).  O-rich irregular LPVs are close to the SRb's lacking a 
      shell and are probably early AGB stars.

\item The difference between L irregular variable stars according to
      their spectral type is evident.  C-rich L stars correspond to
      massive TP-AGB's with a thick envelope.  On the contrary, O-rich
      L stars are close to early AGB O-rich SRb's but their initial
      mass range seems to be more extended to lower masses.

\item One CH star (V Ari) is assigned to the Extended Disk population  
      and it is the faintest star in all the luminosities. This
      completely agrees with the usual hypothesis considering this
      type of star as an old giant star. Its amplitude of variability
      is less than 1 magnitude, a very small value even for a
      semi-regular. The C character of this type of star is
      however difficult to explain. HIPPARCOS observations suggest
      that V Ari could be a suspected  multiple star but this is not
      conclusive.

\end{itemize}

\subsection{Upper limit of the AGB } \label{sec_uplim}

In Section \ref{sec_lumbias} we remarked that the brightest stars in
the sample agree with the brightest luminosity for each group
population (see figure \ref{fig_lumbias}). Thus {\it we can consider
our sample as representative of the LPVs population as far as the
brightest luminosities are concerned}.  Our calibrations show that the
upper limit in K luminosity of the OD population ($(K_0-3\sigma_K) = -8.1$
mag.)  is fainter than that of the D population ($(K_0-3\sigma_K) =
-9.4$ mag.) as seen in Table \ref{tab_estiK}. This confirms the
dependence of the upper limit of the AGB on ${\cal M}_{ms}$. Willson
(1980) has described a schematic evolution on the AGB related to the
mass-loss rate, its acceleration by the pulsations and probably the
induced dust formation.  She found a difference in solar luminosities
of $\sim0.3 L/L_{\sun}$ where stars of solar abundance and ${\cal
M}_{ms}$ equal to 1.5 and 1 ${\cal M}_{\sun}$ leave the AGB.  Our result
is of the same order.

\section{Conclusion} \label{sec_conclu}

Using available HIPPARCOS data we apply the LM algorithm to improve
the luminosity calibrations in visible, near-infrared and infrared
wavelength ranges and to get information about the star and the
circumstellar envelope.  \\

According to the galactic population -- related to initial mass and
metallicity of the stars -- and to the circumstellar envelope
thickness and expansion, several groups of LPVs are obtained: bright
(BD) and disk (disk1)  galactic population with bright and expanding
envelope, not so young and massive disk population (disk2) divided
into 2 groups: one with thin envelope (f) and the other with a bright
and expanding envelope (b). A similar separation according to envelope
properties is found for the old disk (OD) population. At least some
LPVs are found to belong to extended disk (ED)  population.\\

Our results deduced from kinematic properties confirm that the AGB
evolution depends on the initial mass of the progenitor in the main
sequence. This agrees with the comparison of color-magnitude diagrams
using
our estimated individual luminosities with theoretical evolutionary 
tracks.  According to the assigned galactic population we can give
ranges of age and of the lower limit main sequence mass for each star of
our sample. The upper limit of the AGB also depends on ${\cal
M}_{ms}$.   The difference of the values 
 found in K luminosity limits are consistent
with Willson's schematic model related to the mass loss rate and its
acceleration by the pulsations: "Stars evolve up the AGB with only 
moderate mass loss; at $T_e \sim 3000K$ Mira pulsation commences, 
driving the mass loss rate up by at least a factor 10". 
The induced dust formation is followed
by the stabilization of the K luminosity after the carbon enrichment.
\\

The ultimate aim of this work is to estimate individual K, 12 and 25
absolute magnitudes given, in the annex (available as an electronic
table at CDS and in the ASTRID database). This allows us to study 
simultaneously the stellar properties and the 
behavior of the circumstellar envelope.  The results recalled in the
previous paragraph are obtained thanks to the estimated individual
luminosities and they mainly concern properties related to the
assigned galactic populations.  They will be systematically used in
another paper (Mennessier et al., 2001) to study implications regarding the
physics of LPVs, specifically the simultaneous stellar and
circumstellar evolutions along the Asymptotic Giant Branch.\\

\begin{acknowledgements}

This work is supported by the PICASSO program PICS 348 and by the
CICYT under contract ESP97-1803 and AYA2000-0937. We thank A.Gomez for
fruitful discussions of our first results.

\end{acknowledgements}

\end{document}